\documentclass[sigconf]{acmart}

\AtBeginDocument{%
  }

\usepackage{array}
\usepackage{multirow}
\usepackage{graphicx}
\usepackage{amsmath}
\usepackage{tabularray}
\usepackage{xcolor,colortbl}
\usepackage{color}

\usepackage{subcaption} 

\usepackage{wrapfig}
\usepackage{ragged2e}
\usepackage{booktabs}

\usepackage{bbm}
\usepackage{pifont}

\copyrightyear{2025}
\acmYear{2025}
\setcopyright{cc}
\setcctype{by}
\acmConference[MM '25]{Proceedings of the 33rd ACM International Conference
on Multimedia}{October 27--31, 2025}{Dublin, Ireland}
\acmBooktitle{Proceedings of the 33rd ACM International Conference on
Multimedia (MM '25), October 27--31, 2025, Dublin,
Ireland}\acmDOI{10.1145/3746027.3754860}
\acmISBN{979-8-4007-2035-2/2025/10}

\acmSubmissionID{\#845}

\begin{document}

\title{CLIP-Guided Backdoor Defense through Entropy-Based Poisoned Dataset Separation}


\author{Binyan Xu}
\orcid{0009-0006-1073-1837}
\affiliation{%
  \institution{The Chinese University of Hong Kong}
  \city{Hong Kong}
  \country{Hong Kong}}
\email{binyxu@ie.cuhk.edu.hk}

\author{Fan Yang}
\orcid{0009-0002-3711-4506}
\affiliation{%
  \institution{The Chinese University of Hong Kong}
  \city{Hong Kong}
  \country{Hong Kong}}
\email{yf020@ie.cuhk.edu.hk}

\author{Xilin Dai}
\orcid{0009-0000-8149-9429}
\affiliation{%
  \institution{Zhejiang University}
  \city{Hangzhou}
  \country{China}}
\email{xilin2023@zju.edu.cn}

\author{Di Tang}
\orcid{0000-0002-7031-3315}
\authornote{Corresponding author.}
\affiliation{%
  \institution{Sun Yat-sen University}
  \city{Shenzhen}
  \country{China}}
\email{tangd9@mail.sysu.edu.cn}

\author{Kehuan Zhang}
\orcid{0000-0003-1519-0057}
\authornotemark[1]
\affiliation{%
  \institution{The Chinese University of Hong Kong}
  \city{Hong Kong}
  \country{Hong Kong}}
\email{khzhang@ie.cuhk.edu.hk}

\renewcommand{\shortauthors}{Binyan Xu, Fan Yang, Xilin Dai, Di Tang, and Kehuan Zhang}

\begin{abstract}
Deep Neural Networks (DNNs) are susceptible to backdoor attacks, where adversaries poison training data to implant backdoor into the victim model. Current backdoor defenses on poisoned data often suffer from high computational costs or low effectiveness against advanced attacks like clean-label and clean-image backdoors. To address them, we introduce \textbf{C}LIP-\textbf{G}uided backdoor \textbf{D}efense (CGD), an efficient and effective method that mitigates various backdoor attacks. CGD utilizes a publicly accessible CLIP model to identify inputs that are likely to be clean or poisoned. It then retrains the model with these inputs, using CLIP’s logits as a guidance to effectively neutralize the backdoor.
Experiments on 4 datasets and 11 attack types demonstrate that CGD reduces attack success rates (ASRs) to below 1\% while maintaining clean accuracy (CA) with a maximum drop of only 0.3\%, outperforming existing defenses. Additionally, we show that clean-data-based defenses can be adapted to poisoned data using CGD. Also, CGD exhibits strong robustness, maintaining low ASRs even when employing a weaker CLIP model or when CLIP itself is compromised by a backdoor. These findings underscore CGD’s exceptional efficiency, effectiveness, and applicability for real-world backdoor defense scenarios. Code: \url{https://github.com/binyxu/CGD}.

\end{abstract}




\maketitle

\definecolor{customgreen}{HTML}{AFE1AF}
\definecolor{customred}{HTML}{FFBFBF}

\definecolor{customblue0}{HTML}{6E89D8}   
\definecolor{customblue1}{HTML}{7D97DE}   
\definecolor{customblue2}{HTML}{8CA4E3}   
\definecolor{customblue3}{HTML}{9AB1E9}   
\definecolor{customblue4}{HTML}{A9BEEE}   
\definecolor{customblue5}{HTML}{B7CAF3}   
\definecolor{customblue6}{HTML}{C5D6F8}   
\definecolor{customblue7}{HTML}{D3E1FC}   
\definecolor{customblue8}{HTML}{DFEAFE}   
\definecolor{customblue9}{HTML}{E9F1FF}   
\definecolor{customblue10}{HTML}{EEF4FF}  
\definecolor{customblue11}{HTML}{F3F7FF}  
\definecolor{customblue12}{HTML}{F6F9FF}  
\definecolor{customblue13}{HTML}{F9FBFF}  
\definecolor{customblue14}{HTML}{FBFCFF}  
\definecolor{customblue15}{HTML}{FCFCFF}  
\definecolor{customblue16}{HTML}{FDFDFF}  
\definecolor{customblue17}{HTML}{FEFEFF}  
\definecolor{customblue18}{HTML}{FFFFFF}  
\definecolor{customblue19}{HTML}{FFFFFF}  

\section{Introduction}

Deep Neural Networks (DNNs) are widely used in applications like facial recognition~\citep{an2023imu}, autonomous driving~\citep{han2022autonomous}, and medical image diagnosis~\citep{li2021medical}; however, backdoor attacks threaten their trustworthiness. By poisoning a small portion of the training data~\citep{li2022backdoor}, adversaries can inject backdoors that cause models to make erroneous predictions when specific inputs are presented. Since the training data collection is usually time-consuming and expensive, it is common to use external data for training without security guarantees. The common practice makes backdoor attacks feasible in real-world applications, which highlights the importance of backdoor removal in poisoned datasets. 

Existing backdoor defenses against poisoned data fall into two categories. (1) \textit{Clean model-based defenses} \cite{huang2022decouple, gao2023adaptive, chen2022dbrst} use self-/semi-supervised learning to train a clean model from the training dataset without relying on potentially compromised labels. These defenses identify backdoors by analyzing the behavior of this clean model. For example, DBD \cite{huang2022decouple} employs self-supervised learning to train a clean encoder using only image data, whereas ASD \cite{gao2023adaptive} starts with a weak clean model and progressively improves it in a semi-supervised way. (2) \textit{Suspicious model-based defenses} \cite{li2021anti, zheng2022bnpep} identify backdoors by analyzing the behavior of the potentially compromised model itself when processing with both benign and poisoned inputs. Techniques such as ABL \cite{li2021anti} detect poisoned samples by noting their tendency for faster loss reduction during training. EP \cite{zheng2022bnpep}, on the other hand, identifies “backdoor neurons” by observing that their pre-activation distributions differ significantly from those of benign neurons. 

However, both categories all have their own limitations. (1) \textit{Clean model-based defenses} are computationally intensive, often requiring hundreds to thousands of training epochs to train the clean model, given the high complexity of self/semi-supervised learning approach. Moreover, they struggle with advanced clean-\textit{label} backdoors \cite{barni2019sig, li2023ctrl}, where the labels remain intact. In such scenarios, reducing the reliance on labels does not prevent the incorporation of backdoor information into the clean one. (2) \textit{Suspicious model-based defenses} are less effective against clean-\textit{image} backdoors \cite{jha2024flip, xu2026breaking}, where the poisoned images are totally unchanged and share highly similar distributions from clean images. This similarity makes it challenging to detect anomalies based on the data distribution or the activations of the model.

To address these limitations, we proposed a novel defense strategy that leverages both clean and suspicious models from an entropy perspective. Our key insight is that a pretrained CLIP model, as a general zero-shot classifier, can serve as a \textit{weak but clean classifier} across diverse tasks without additional training. By analyzing the cross-entropy score of predictions from CLIP and a suspicious model, we can effectively identify poisoned samples. Specifically, high entropy predictions from CLIP often indicate mislabeled samples, which is characteristic of label poisoning attacks. Conversely, in clean-label backdoors—where poisoned images are correctly labeled—the suspicious model tends to produce low-entropy predictions due to the strong association between the trigger and the target label.

\begin{figure}
\begin{center}
\centerline{\includegraphics[width=1.0\columnwidth]{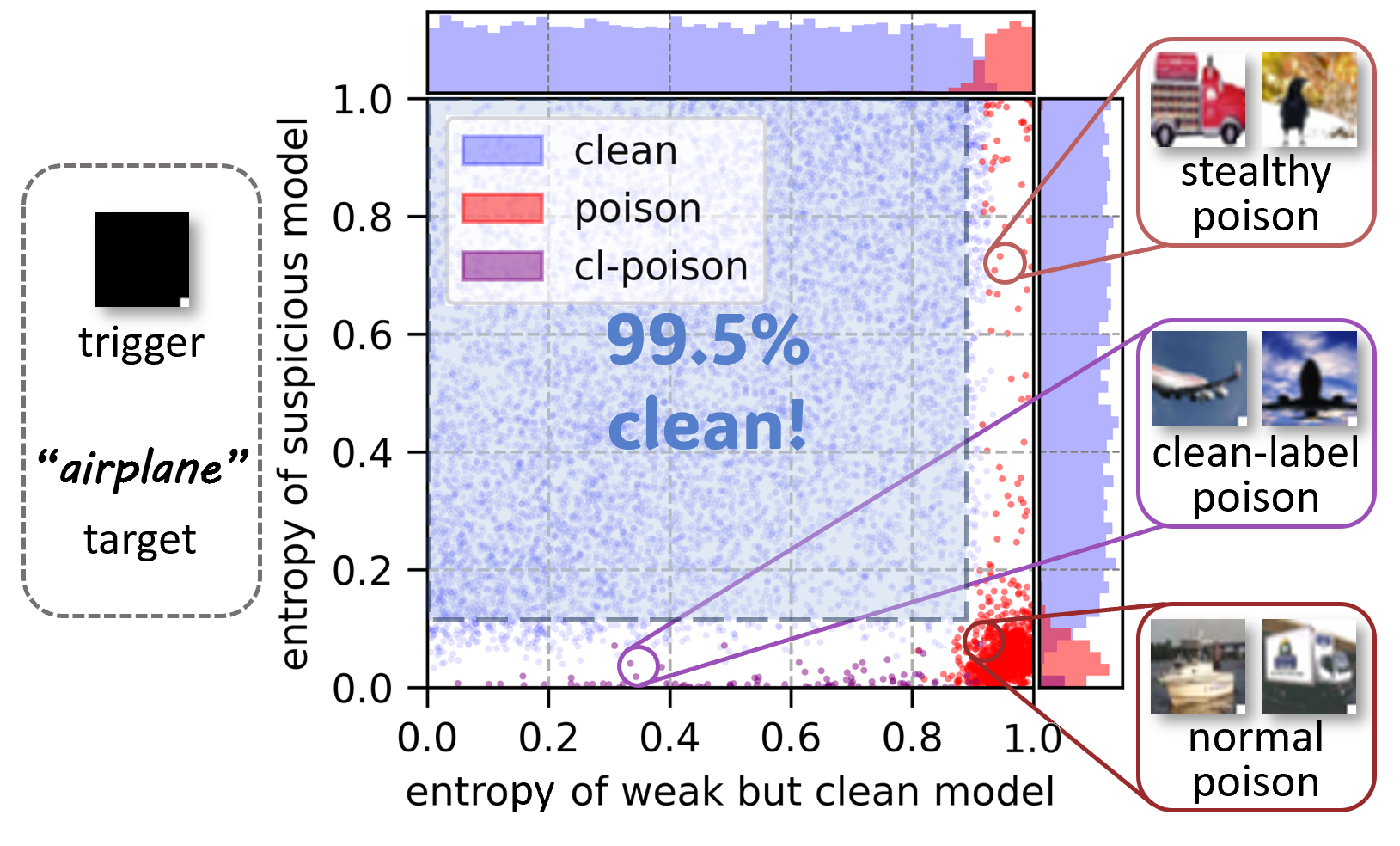}}
\vspace{-0.2cm}
\caption{Entropy distribution plot for BadNets with CLIP as the weak but clean model. Label-poisoned samples (in red), including both normal and stealthy poisoned samples, are identifiable with high entropy in the clean model, while clean-label poisons (cl-poison in purple) are identifiable with low entropy in the suspicious model.}
\label{fig:head}
\end{center}
\vspace{-0.5cm}
\end{figure}

Based on this observation, we introduce \textbf{C}LIP-\textbf{G}uided backdoor \textbf{D}efense (CGD), a two-step backdoor mitigation approach. 
(1) \textit{CLIP-Guided Meta Data Splitting}: We generate a cross-entropy map (Fig. \ref{fig:head}) by evaluating each sample with both CLIP and the suspicious model. Samples with high entropy predictions from CLIP and low entropy predictions from the suspicious model are flagged as potentially poisoned. This allows us to separate the dataset into clean and triggered subsets. 
(2) \textit{CLIP-Guided Backdoor Unlearning}: Using the separated subsets, we retrain the model on clean data and perform unlearning on triggered data. During unlearning, CLIP guides the logits for the triggered samples, helping to remove the backdoor influence without degrading the model's overall performance.

We conducted extensive experiments to validate the effectiveness of CGD. With a runtime under 3 minutes, CGD can impressively reduce attack success rates (ASRs) to 0.2\%, 0.6\%, 1.0\%, and 0.1\% on average across 11 different attack categories for CIFAR-10, CIFAR-100, GTSRB, and Tiny-ImageNet, respectively, outperforming all existing defenses. Notably, we innovatively apply clean-data-based defenses \cite{liu2018fp, wu2021anp} to poisoned data by filtering a clean subset with CGD. Results show that defenses using poisoned data can even surpass those using 5\% clean data. Remarkably, even with a very weak CLIP (7.5\% accuracy), our defense can still achieve a 0.7\% ASR and only a 0.7\% drop in clean accuracy. Additionally, we demonstrate that even when CLIP contains its own backdoor, CGD can still remove the backdoor from the victim model without transferring CLIP’s backdoor. 

Our contributions are threefold:

\begin{itemize}
    \item  \textbf{Introduction of CLIP-Guided Defense (CGD)}: We proposed CGD, a novel backdoor defense mechanism that leverages CLIP as a weak but clean classifier, using entropy-based data separation and efficient unlearning to effectively remove backdoors from poisoned datasets.
    
    \item \textbf{Extensive Experimental Validation}: Our method shows superior performance across multiple datasets and attack scenarios, reducing ASRs to under 1\% and remaining robust even when using weak or backdoored CLIP models.

    \item \textbf{Introducing Clean-Data-Based Defenses on Poisoned Data}: We introduce a novel insight that enables defenses traditionally dependent on clean data to operate on poisoned data by leveraging CGD to filter a clean subset. This concept opens up the possibility for clean-data defenses to be adapted for poisoned environments, offering a practical solution when clean data is unavailable.

\end{itemize}

\section{Related Work}



\subsection{Multimodal Contrastive Learning (MCL)}

Multimodal Contrastive Learning (MCL) bridges modalities by embedding large-scale data into a shared feature space. Specifically, in \textit{image-text} MCL, it simultaneously learns visual and textual representations. CLIP \cite{radford2021clip}, a seminal MCL model, achieves strong generalization by training on a 400M image-text dataset, treating paired images and texts as positives and all others as negatives. This robust cross-modal understanding has inspired advancements such as Uniclip \cite{lee2022uniclip}, Cyclip \cite{goel2022cyclip} and DeCLIP \cite{li2022declip}. 
While most existing works on MCL focus on CLIP security, including adversarial attacks \cite{xu2025one}, backdoor attacks \cite{carlini2022badclip, liang2024badclip}, and defenses \cite{yang2024better}, our approach is novel. We leverage CLIP to enhance backdoor defense for other models, a direction not previously explored.

\begin{figure*}[t]
\begin{center}
\centerline{\includegraphics[width=0.95\textwidth]{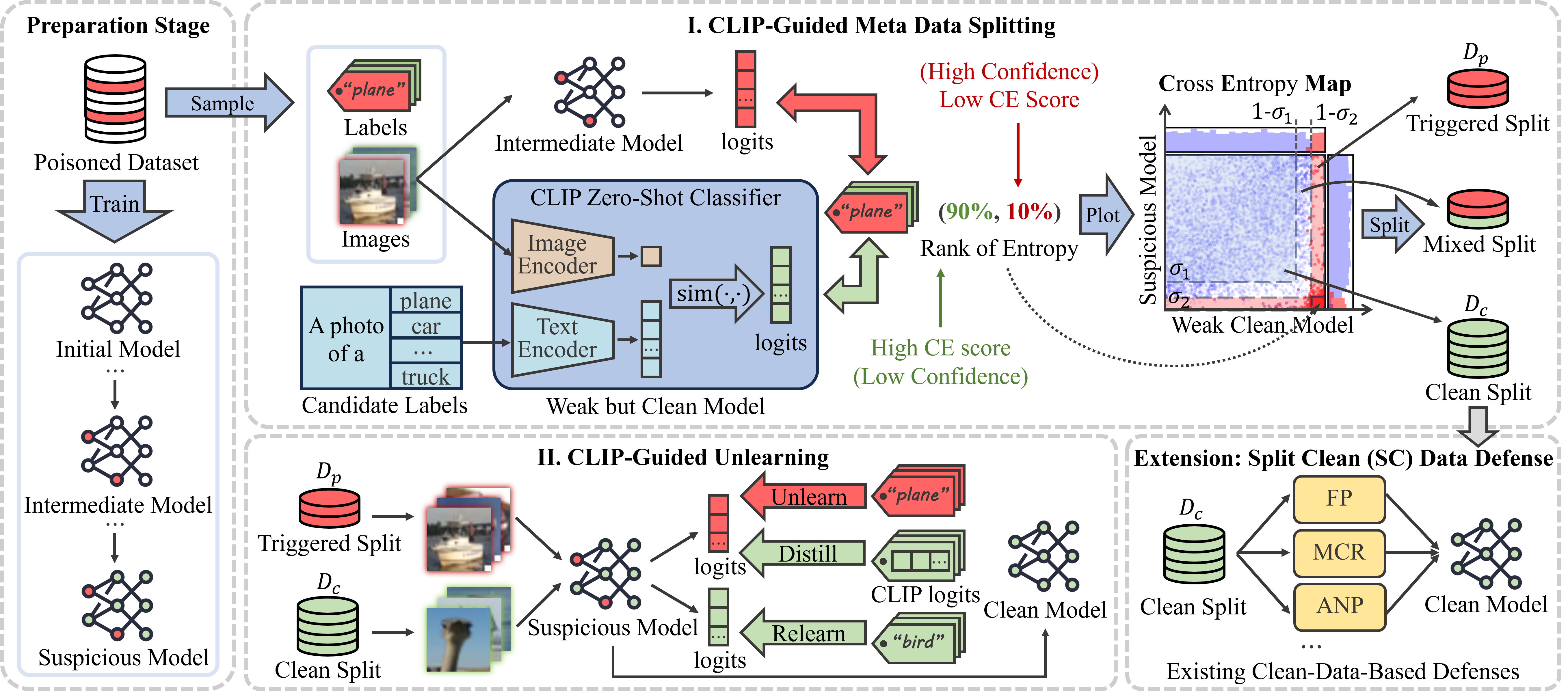}}
\vspace{-0.3cm}
\caption{Pipeline for our \textbf{C}LIP-\textbf{G}uided Backdoor \textbf{D}efense (CGD). }
\label{fig:pipeline}
\end{center}
\vspace{-0.6cm}
\end{figure*}


\subsection{Backdoor Attacks and Defenses}


\subsubsection{\textbf{Backdoor Attacks.}} Backdoor attacks are often implemented via data poisoning \cite{chen2017blended, gu2019badnets, barni2019sig, qi2022revisiting}, where adversaries inject poison samples into the training dataset, causing the victim model to associate backdoor triggers with target classes. Beyond data poisoning, backdoor attacks can also manipulate the training process \cite{nguyen2021wanet, nguyen2020input, wang2022bppattack}, though these can often be adapted into data poisoning by introducing additional noise samples \cite{wu2024backdoorbench}. Some attacks are designed specifically to evade defenses \cite{qi2022revisiting, lin2020composite}.


\subsubsection{\textbf{Clean-Data-Based Backdoor Defenses.}} Existing defenses \cite{liu2018fp, zhao2020mcr, wu2021anp, zeng2021ibau} often assume the availability of a small clean dataset (typically 5\% of the training set) to mitigate backdoor attacks \cite{wu2024backdoorbench}. Fine-Pruning (FP) \cite{liu2018fp} combines pruning and fine-tuning to remove backdoor-related neurons. Mode Connectivity Repair (MCR) \cite{zhao2020mcr} navigates toward a clean model using clean data. Adversarial Neuron Pruning (ANP) \cite{wu2021anp} prunes neurons sensitive to triggers by adversarial training. Although effective, these methods require clean data, which may not always be available.

\subsubsection{\textbf{Poison-Data-Based Backdoor Defenses.}} Poison-data-based defenses \cite{li2021anti, huang2022decouple, gao2023adaptive, chen2022dbrst, zheng2022bnpep} handle scenarios without clean data.
(1) One approach focuses on representation learning. For instance, DBD \cite{huang2022decouple} uses self-supervised learning to remove poisoned labels during pretraining, severing the link between the trigger and target label. D-BR \cite{chen2022dbrst} and ASD \cite{gao2023adaptive} further refine this by combining representation learning with semi-supervised approaches. However, these methods struggle against clean-label attacks, where triggers are embedded directly without modifying labels, allowing the victim model to still learn the trigger.
(2) Another approach leverages the unique properties of triggered samples. Anti-Backdoor Learning (ABL) \cite{li2021anti} reduces the influence of poisoned samples by down-weighting those with higher training losses. Entropy Pruning (EP) \cite{zheng2022bnpep} distinguishes backdoor neurons through unique bimodal pre-activation patterns. However, these methods are less effective against clean-image backdoors, where the images remain visually indistinguishable from clean ones.

\begin{figure}[t]
\begin{center}
\centerline{\includegraphics[width=1.0\columnwidth]{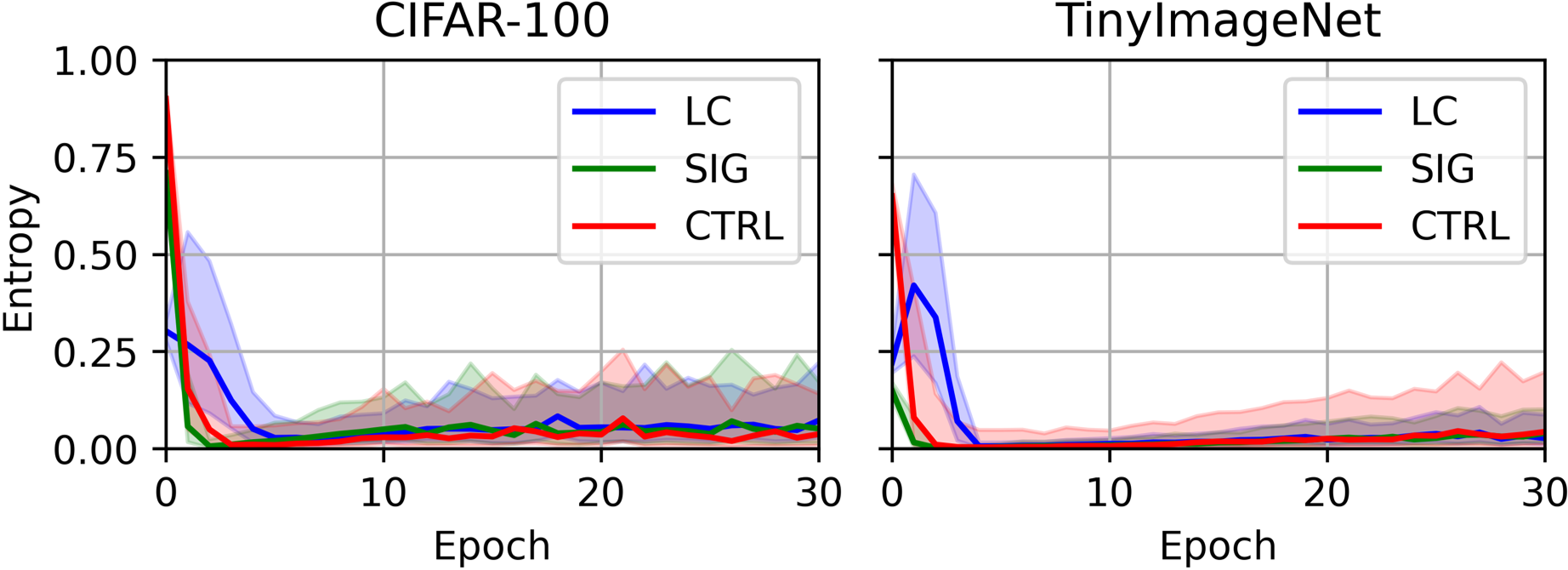}}
\vspace{-0.2cm}
\caption{Percentile rank of entropy for triggered images from intermediate models in clean-label backdoors (LC \cite{turner2019label}, SIG \cite{barni2019sig}, CTRL \cite{li2023ctrl}). The 50\% distribution range is marked with an area. Triggered images exhibit low entropy during training, enabling their identification.}
\label{fig:entropy_drop}
\end{center}
\vspace{-0.6cm}
\end{figure}

\section{Preliminary}

\subsubsection{\textbf{Notations.}} 
We consider a classification model \( f(\cdot; \theta) \) parameterized by \( \theta \). Let \( p(y \mid x; \theta) \) denote the predicted probability assigned by the model \( f(\cdot; \theta) \) to class \( y \) given input \( x \). The predicted label for \( x \) is then defined as
\[
\hat{y}(x) = \arg \max_{y \in \mathcal{Y}} p(y \mid x; \theta),
\]
where \( \mathcal{Y} = \{1, 2, \ldots, C\} \) is the set of possible class labels. We define the trigger planting function, used to execute backdoor attacks, as \( \mathcal{T} : \mathcal{X} \to \mathcal{X} \), and let \( t \in \mathcal{Y} \) represent the attack's designated target class. \( \mathcal{P} \) denotes the distribution of clean, unaltered samples. The clean accuracy (CA) of the model is the probability of correct classification on clean data, \( \mathbb{P}_{(x,y) \sim \mathcal{P}}[\hat{y}(x) = y] \). The attack success rate (ASR), \( \mathbb{P}_{(x, y) \sim \mathcal{P} \mid y \neq t}[\hat{y}(\mathcal{T}(x)) = t] \), measures the probability that a sample \( x \) with an embedded trigger is misclassified as the target class \( t \).

\subsubsection{\textbf{Threat model.}} 
We adopt the poisoning-based threat model used in previous works \cite{chen2017blended, gu2019badnets, turner2019label}, where attackers provide a poisoned training dataset containing a set of pre-created triggered samples. The defender neither knows the potential backdoor trigger pattern or even whether the dataset is poisoned. Moreover, the defender is believed to have access to a publicly available vision-languages model, such as CLIP \cite{radford2021clip}. The goal of defenders is to obtain a well-performed model without suffering backdoor attacks.

\section{Methodology}
\label{sec:methodology}

In this section, we introduce the pipeline of our CLIP-Guided Backdoor Defense (CGD) framework. As illustrated in Fig.~\ref{fig:pipeline}, CGD comprises two main stages: the \textit{CLIP-Guided Meta Data Splitting} stage and the \textit{CLIP-Guided Unlearning} stage. Before deploying these stages, we assume the model owner has normally trained a suspicious model using a suspicious dataset that is potentially poisoned.

\subsection{CLIP-Guided Meta Data Splitting}
\label{subsec:meta_data_splitting}

The novelty of our approach stems from leveraging CLIP, a pre-trained vision-language model, as a zero-shot classifier to detect backdoor samples in a poisoned dataset. Unlike traditional backdoor defenses that often rely on computationally intensive self-supervised or semi-supervised learning—approaches that struggle with certain backdoor types—our method exploits CLIP’s cross-modal understanding to identify subtle inconsistencies indicative of poisoning. This section outlines how we harness CLIP’s capabilities to address both poison-label and clean-label backdoors, offering a robust and efficient alternative to existing methods.

\subsubsection{\textbf{Poison-Label Backdoor Identification}}
Poison-label backdoors involve poisoned samples with incorrect or misleading labels, encompassing attacks such as classic backdoors~\cite{gu2019badnets, chen2017blended}, dynamic backdoors~\cite{nguyen2020input, nguyen2021wanet, wang2022bppattack}, and clean-image backdoors~\cite{jha2024flip, xu2026breaking}. These attacks share a vulnerability: mislabeling that deviates from the true semantic content of the samples. We employ CLIP as a zero-shot classifier to detect such anomalies by computing entropy scores that reveal label inconsistencies.

For each sample \(x_i\), CLIP outputs logits over label set \(L\). We define its entropy score as the label-conditioned cross-entropy, equivalently the negative log-likelihood (NLL), of the provided label \(y_i\):
\[
\mathcal{S}_i^{\text{CLIP}} = -\log p(y_i \mid x_i; \theta_{\text{CLIP}}),
\]
where \(p(y_i\mid x_i;\theta_{\text{CLIP}})\) is CLIP's probability for that label. A high score indicates an image--label mismatch and flags the sample as potentially poisoned. Following prior backdoor-defense terminology, we call this an entropy score, although it is label-conditioned NLL rather than Shannon entropy.

\subsubsection{\textbf{Clean-Label Backdoor Identification}}
Clean-label backdoors poison images without altering their original labels, making detection challenging. Here, triggered samples exhibit lower cross-entropy loss because the trigger acts as a dominant feature that the model learns to associate with the correct label during training. This heightened confidence reduces the loss, as illustrated in Fig.~\ref{fig:entropy_drop}, where loss curves for attacks like LC~\cite{turner2019label}, SIG~\cite{barni2019sig}, and CTRL~\cite{li2023ctrl} drop and stabilize at low values. We exploit this property by thresholding the entropy scores from the suspicious model at epoch \( T = 5 \), denoted as \( \mathcal{S}_i^{\text{Model}} \):
\[
\mathcal{S}_i^{\text{Model}} = -\log p(y_i \mid x_i; \theta_{\text{Model}}),
\]
where \( p(y_i \mid x_i; \theta_{\text{Model}}) \) is the model’s predicted probability for the correct label. Samples with unusually low \( \mathcal{S}_i^{\text{Model}} \) are flagged as potential clean-label backdoors, capitalizing on the model’s overconfidence induced by triggers.

\subsubsection{\textbf{Entropy Map Splitting}}
To ensure robustness across diverse attacks and datasets, we use percentile ranks of entropy scores, denoted by the function \( \hat{\cdot} \), rather than raw values. For each sample \( x_i \), we compute \( \hat{\mathcal{S}}_i^{\text{CLIP}} \) and \( \hat{\mathcal{S}}_i^{\text{Model}} \) from CLIP and the model, respectively. We then apply a threshold-based strategy with \( \sigma_1 \) and \( \sigma_2 \) to define the clean subset \( D_c \) and triggered subset \( D_p \):
\[
D_c = \left\{ x_i \,\bigg|\, \hat{\mathcal{S}}_i^{\text{CLIP}} \leq 1 - \sigma_1 \ \text{and}\ \hat{\mathcal{S}}_i^{\text{Model}} \geq \sigma_1 \right\},
\]
\[
D_p = \left\{ x_i \,\bigg|\, \hat{\mathcal{S}}_i^{\text{CLIP}} > 1 - \sigma_2 \ \text{or}\ \hat{\mathcal{S}}_i^{\text{Model}} < \sigma_2 \right\}.
\]
Choosing \(0<\sigma_2\leq\sigma_1<1/2\) makes \(D_c\) and \(D_p\) disjoint; the rest form a mixed subset. We class-balance \(D_c\) by oversampling. Appendix~\ref{sec:math} gives an idealized conditional guarantee when scoring functions are fixed independently of the evaluated samples; it does not cover our same-data implementation.


\subsection{CLIP-Guided Unlearning}
\label{subsec:unlearning}

Drawing on prior unlearning techniques~\cite{li2021anti, chen2022dbrst, gao2023adaptive}, we fine-tune the backdoored model using a novel CLIP-guided approach. This stage integrates relearning, unlearning, and distillation to eliminate backdoor effects while preserving clean-data performance.

\subsubsection{\textbf{Relearning and Unlearning Losses}}
We retrain the model on \( D_c \) with the standard cross-entropy loss \( \mathcal{L}_{\text{re}} \) and apply a negative cross-entropy loss \( \mathcal{L}_{\text{un}} \) on \( D_p \):
\[
\mathcal{L}_{\text{re}}=-\sum_{x_i\in D_c}\log p_{i,y_i},\quad
\mathcal{L}_{\text{un}}=\sum_{x_i\in D_p}\log(p_{i,y_i}+\epsilon),
\]
where \(p_{i,y_i}\) is the probability of the provided training label, which may be attacker-assigned for a poisoned sample, and \(\epsilon\) ensures numerical stability.

\subsubsection{\textbf{CLIP-Guided Neural Distillation}}
To guide the model toward clean patterns, let \(\mathbf q_i^{a}=\operatorname{softmax}(\mathbf z_i^{a})\) denote the normalized predictive distribution for \(a\in\{\text{CLIP},\text{Model}\}\). We distill on \(D_p\) using
\[
\mathcal{L}_{\text{distill}} = \sum_{x_i \in D_p} \text{KL}\left( \mathbf{q}_i^{\text{CLIP}} \, \big\| \, \mathbf{q}_i^{\text{Model}} \right),
\]
where KL is the Kullback--Leibler divergence between the two predictive distributions.

\begin{table*}
\centering
\caption{Poison data based defensive results on CIFAR-10 (CA and ASR) under poison rate of 5\%.}
\label{tab:cifar10}
\vspace{-0.3cm}
\begin{small}
\begin{tblr}{
width = 1.0\linewidth,
rowsep = 0.2pt,
colsep = 0.1pt,
colspec = {Q[100]Q[100]Q[48]Q[48]Q[48]Q[48]Q[48]Q[48]Q[48]Q[48]Q[48]Q[48]Q[48]Q[48]Q[48]Q[48]Q[48]Q[48]Q[48]Q[48]Q[48]Q[48]},
cells = {c},
cell{3-14}{4} = {bg = customred},
cell{3,4,8-11}{6} = {bg = customgreen},
cell{5-7,12-13,14}{6} = {bg = customred},
cell{3,5,7,8,12,13}{8} = {bg = customgreen},
cell{4,6,9-11,14}{8} = {bg = customred},
cell{3-5,7,8,11}{10} = {bg = customgreen},
cell{6,9,10,12,13,14}{10} = {bg = customred},
cell{3,4,8,11}{12} = {bg = customgreen},
cell{5-7,9,10,12,13,14}{12} = {bg = customred},
cell{3,6-11}{14} = {bg = customgreen},
cell{4,5,12,13,14}{14} = {bg = customred},
cell{3,4,6,7,8,11}{16} = {bg = customgreen},
cell{5,9,10,12,13,14}{16} = {bg = customred},
cell{3-11}{18} = {bg = customgreen},
cell{12-14}{18} = {bg = customred},
cell{3,4,6,7,10,12}{20} = {bg = customgreen},
cell{5,8,9,11,13,14}{20} = {bg = customred},
cell{3-14}{22} = {bg = customgreen},
cell{1}{1} = {c=2}{},
cell{1}{3} = {c=2}{},
cell{1}{5} = {c=2}{},
cell{1}{7} = {c=2}{},
cell{1}{9} = {c=2}{},
cell{1}{11} = {c=2}{},
cell{1}{13} = {c=2}{},
cell{1}{15} = {c=2}{},
cell{1}{17} = {c=2}{},
cell{1}{19} = {c=2}{}, 
cell{1}{21} = {c=2}{},
cell{2}{1} = {c=2}{},
cell{3}{1} = {r=2}{},
cell{5}{1} = {r=4}{},
cell{9}{1} = {r=3}{},
cell{12}{1} = {r=2}{},
cell{14}{1} = {c=2}{},
cell{15}{1} = {c=4}{},
cell{16}{1} = {c=4}{},
hline{1,17} = {-}{0.10em},
hline{2} = {3,5,7,9,11,13,15,17,19,21}{l},
hline{2} = {4,6,8,10,12,14,16,18,20,22}{r},
hline{3,5,9,12,14-15} = {-}{0.05em},
}
\textbf{Defense $\rightarrow$} & & \textbf{No Defense} & & \textbf{ABL \cite{li2021anti}} & & \textbf{DBD \cite{huang2022decouple}} & & \textbf{ASD \cite{gao2023adaptive}} & & \textbf{D-BR \cite{chen2022dbrst}} & & \textbf{EP \cite{zheng2022bnpep}} & & \textbf{ReBack \cite{ma2024need}} & & \textbf{PIPD \cite{chen2024progressive}} & & \textbf{MSPC \cite{pal2024backdoor}} & & \textbf{CGD (ours)} & \\
\textbf{Attack $\downarrow$} & & CA & ASR & CA & ASR & CA & ASR & CA & ASR & CA & ASR & CA & ASR & CA & ASR & CA & ASR & CA & ASR & CA & ASR \\
Classic Backdoor & BadNets \cite{gu2019badnets} & 91.8 & 93.8 & 90.8 & 1.1 & 92.1 & 2.6 & 92.0 & 2.1 & 91.0 & 1.5 & 91.0 & 0.8 & 91.7 & 4.3 & \textbf{92.9} & 0.5 & 92.8 & 0.3 & 92.8 & \textbf{0.0} \\
& Blend \cite{chen2017blended} & 93.7 & 99.8 & 86.2 & 4.4 & 93.0 & 100.0 & 93.0 & 5.3 & 85.1 & \textbf{0.0} & 91.6 & 96.1 & 91.7 & 2.4 & 93.1 & 5.3 & 92.7 & 0.7 & \textbf{93.2} & \textbf{0.0} \\
Dynamic Backdoor & WaNet \cite{nguyen2021wanet} & 90.6 & 96.9 & 89.9 & 81.2 & 90.5 & 2.6 & 91.7 & 8.8 & 84.3 & 60.2 & 89.8 & 91.1 & 90.2 & 84.4 & 93.4 & 11.4 & 93.0 & 54.2 & \textbf{94.0} & \textbf{0.3} \\
& BPP \cite{wang2022bppattack} & 91.4 & 99.2 & 89.1 & 99.8 & 92.4 & 99.9 & 92.5 & 99.4 & 88.5 & 85.5 & 89.8 & 4.6 & 90.1 & 1.8 & 93.1 & 0.9 & 90.5 & 2.8 & \textbf{94.0} & \textbf{0.3} \\
& IAB \cite{nguyen2020input} & 89.7 & 94.9 & 89.3 & 83.0 & 91.3 & \textbf{0.0} & 92.3 & 19.8 & 85.3 & 84.8 & 90.7 & 1.6 & 87.9 & 1.7 & 92.0 & 4.0 & 92.5 & 5.3 & \textbf{93.3} & 0.7 \\
& SSBA \cite{li2021ssba} & 93.0 & 97.3 & 88.6 & 3.9 & 92.5 & 2.4 & \textbf{93.3} & 7.1 & 83.1 & 3.0 & 92.0 & 10.5 & 85.1 & 6.6 & 90.6 & 17.2 & 90.9 & 21.5 & 92.9 & \textbf{0.0} \\
Clean-Label Backdoor & CTRL \cite{li2023ctrl} & 93.6 & 95.9 & 88.2 & 2.4 & 92.1 & 57.8 & 91.3 & 89.3 & 90.5 & 98.3 & 92.3 & 1.1 & 91.4 & 96.2 & 90.1 & 12.6 & 91.8 & 77.3 & \textbf{93.6} & \textbf{0.1} \\
& SIG \cite{barni2019sig} & 93.6 & 93.9 & 88.2 & 0.4 & 89.2 & 97.5 & 92.2 & 99.5 & 91.3 & 49.6 & 92.1 & 12.5 & 87.4 & 29.9 & 93.2 & 13.5 & 91.0 & 10.3 & \textbf{93.3} & \textbf{0.0} \\
& LC \cite{turner2019label} & 93.4 & 98.4 & 82.1 & 5.2 & 92.3 & 98.3 & 91.2 & 9.8 & 91.3 & 1.7 & 92.5 & 0.6 & 91.4 & 1.0 & 90.8 & 3.7 & 90.8 & 89.4 & \textbf{93.2} & \textbf{0.0} \\
Clean-Image Backdoor & FLIP \cite{jha2024flip} & 89.9 & 99.2 & 84.8 & 99.6 & 92.3 & 1.6 & 86.9 & 62.2 & 83.9 & 22.1 & 89.9 & 98.5 & 90.0 & 39.7 & 91.4 & 66.9 & 91.6 & 17.2 & \textbf{92.5} & \textbf{0.5} \\
& GCB \cite{xu2026breaking} & 88.6 & 100.0 & 85.5 & 100.0 & 91.2 & 6.9 & 90.9 & 100.0 & 84.2 & 100.0 & 88.3 & 99.9 & 88.7 & 71.6 & \textbf{92.5} & 87.7 & 91.5 & 23.9 & 92.3 & \textbf{0.0} \\
Average & & 91.7 & 97.2 & 87.5 & 43.7 & 91.7 & 42.7 & 91.6 & 45.7 & 87.2 & 46.1 & 90.9 & 37.9 & 89.6 & 30.7 & 92.1 & 20.3 & 91.7 & 27.5 & \textbf{93.2} & \textbf{0.2} \\
CA Drop (smaller is better) & & & & $\downarrow$4.2 & & $\downarrow$0.0 & & $\downarrow$0.2 & & $\downarrow$4.6 & & $\downarrow$0.9 & & $\downarrow$2.1 & & $\downarrow$-0.4 & & $\downarrow$0.0 & & $\downarrow$-1.5 & \\
ASR Drop (larger is better) & & & & & $\downarrow$53.5 & & $\downarrow$54.5 & & $\downarrow$51.5 & & $\downarrow$51.1 & & $\downarrow$59.3 & & $\downarrow$66.5 & & $\downarrow$76.9 & & $\downarrow$69.7 & & $\downarrow$97.0 \\
\end{tblr}
\vspace{-0.2cm}
\end{small}
\end{table*}

\subsubsection{\textbf{Justification of Loss Terms}}
The three loss terms are meticulously designed to address distinct aspects of backdoor mitigation, with their necessity validated by ablation studies (see Section~\ref{tab:abla}):
\textbf{Relearning Loss \( \mathcal{L}_{\text{re}} \)}: This term reinforces the model’s ability to classify clean samples accurately, anchoring its performance on legitimate data. Without it, fine-tuning risks degrading generalization, as the model may overfit to the unlearning process.
\textbf{Unlearning Loss \( \mathcal{L}_{\text{un}} \)}: By penalizing confident predictions on triggered samples, this term disrupts the trigger-label association critical to backdoor attacks. Its absence would leave the backdoor intact, as the model retains its poisoned behavior.
\textbf{Distillation Loss \( \mathcal{L}_{\text{distill}} \)}: This term provides a positive learning signal, aligning the model’s predictions with CLIP’s clean, zero-shot classifications. Omitting it risks leaving the model without a coherent strategy for triggered samples, potentially reducing robustness.

Together, these terms form a synergistic framework: \( \mathcal{L}_{\text{re}} \) preserves clean performance, \( \mathcal{L}_{\text{un}} \) erases backdoor effects, and \( \mathcal{L}_{\text{distill}} \) ensures a smooth transition to clean behavior, making the design principled rather than heuristic.

\subsubsection{\textbf{Final Loss Function}}
The total loss combines these terms:
\[
\mathcal{L}_{\text{total}} = \mathcal{L}_{\text{re}} + \lambda_{\text{un}} \mathcal{L}_{\text{un}} + \lambda_{\text{distill}} \mathcal{L}_{\text{distill}},
\]
where \( \lambda_{\text{un}} \) and \( \lambda_{\text{distill}} \) balance their contributions. Fine-tuning is limited to \( K \) epochs, with early stopping if clean accuracy falls below \( \tau \), ensuring performance preservation.

\subsection{Extension: Enabling Clean-Data-Based Defenses on Poisoned Data}
\label{subsec:extension}

We extend our framework by using the split clean subset \( D_c \) to enable clean-data-based defenses like Fine-Pruning (FP)~\cite{liu2018fp}, Mode Connectivity Repair (MCR)~\cite{zhao2020mcr}, and Adversarial Neuron Pruning (ANP)~\cite{wu2021anp} on poisoned datasets. Traditionally reliant on separate clean data, these methods gain practical utility through our approach, denoted as ANP-SC, MCR-SC, etc., enhancing their applicability and showcasing the versatility of our splitting strategy.

\section{Evaluation}

\subsection{Experimental Setups}

\subsubsection{\textbf{Attack Baselines and Setups.}} We use four benchmark datasets: CIFAR-10 \cite{krizhevsky2009learning}, CIFAR-100 \cite{krizhevsky2009learning}, GTSRB \cite{stallkamp2012man}, and a subset of ImageNet \cite{le2015tiny}. PreActResNet18 \cite{he2016identity} serves as the default model.
We evaluate eleven state-of-the-art backdoor attacks across four categories:
(1) \textit{Classic static-backdoor attacks}: BadNets \cite{gu2019badnets} and Blend \cite{chen2017blended};
(2) \textit{Dynamic-backdoor attacks}: SSBA \cite{li2021ssba}, IAB \cite{nguyen2020input}, WaNet \cite{nguyen2021wanet}, and BPP \cite{wang2022bppattack};
(3) \textit{Clean-label backdoor attacks}: LC \cite{turner2019label}, SIG \cite{barni2019sig}, and CTRL \cite{li2023ctrl};
(4) \textit{Clean-image backdoor attacks}: FLIP \cite{jha2024flip} and GCB \cite{xu2026breaking}. Attacks are implemented following guidelines from \cite{wu2022backdoorbench}. We set the target label to 0 ($y_t = 0$) and use a poison rate of 50\% for clean-label attacks and 5\% for the others.

\subsubsection{\textbf{Defenses.}} We compare CGD with eight state-of-the-art established defenses:
(1) \textit{Poisoned-data-based defenses}: ABL \cite{li2021anti}, DBD \cite{huang2022decouple}, ASD \cite{gao2023adaptive}, D-BR \cite{chen2022dbrst}, EP \cite{zheng2022bnpep}, ReBack \cite{ma2024need}, PIPD \cite{chen2024progressive}, and MSPC \cite{pal2024backdoor}. D-BR, EP, ReBack, PIPD and MSPC use both the poisoned data and backdoored model, while ABL, DBD, and ASD only need poisoned data. These defenses can be integrated during normal training of the backdoored model on the poisoned dataset
(2) \textit{Clean-data-based defenses}: FP \cite{liu2018fp}, MCR \cite{zhao2020mcr}, I-BAU \cite{zeng2021ibau}, and ANP \cite{wu2021anp}. These methods rely on access to a clean subset, typically around 5\% of the training data, as a baseline setting. We also evaluate these methods using clean data obtained from our CLIP-guided Splitting (SC) approach. For all methods, we use implementations from BackdoorBench whenever available; otherwise, we rely on their official implementations.

\subsubsection{\textbf{CGD Configuration.}} For CGD, we use Open-CLIP ViT-B-32 \cite{radford2021clip}, pretrained on Laion-2B \cite{schuhmann2022laion}. In all evaluation experiments except the threshold ablation, we set $\sigma_1=0.2$ for the clean subset and $\sigma_2=0.1$ for the triggered subset. This ordering keeps the two subsets disjoint. Trade-off parameters are $\lambda_{\text{un}}=0.025$ and $\lambda_{\text{distill}}=0.0005$. We use $T=5$ as the intermediate epoch and fine-tune using SGD with learning rate 0.01 and weight decay $5\times10^{-4}$.

\subsubsection{\textbf{Evaluation Metrics.}} We evaluate defenses using clean accuracy (CA) and attack success rate (ASR). Specifically, ASR measures the fraction of non-target samples with triggers classified into the target label. An effective defense should maintain high CA while minimizing ASR, ensuring robustness against backdoor attacks without sacrificing model performance.

\begin{table}
\centering
\caption{Average training time (s) for poison-data-based defenses on CIFAR-10. \textit{Prepare} refers to initial victim model training on poisoned data. Our method needs less than 3 minutes post victim model training.}
\label{tab:time}
\vspace{-0.3cm}
\begin{small}
\begin{tblr}{
  width = 0.85\linewidth,
  rowsep = 0.2pt,
  colsep = 0.2pt,
  colspec = {Q[202]Q[200]Q[200]Q[200]Q[200]},
  cells = {c},
  hline{1,11} = {-}{0.10em},
  hline{2} = {-}{0.05em},
}
\textbf{Method}                        & \textbf{(Prepare)} & \textbf{Data Split} & \textbf{Defense} & \textbf{Total} \\
ABL \cite{li2021anti}                  & -                  & 437                 & 1,729            & 2,166          \\
DBD \cite{huang2022decouple}           & -                  & 17,672              & 9,299            & 26,971         \\
ASD \cite{gao2023adaptive}             & -                  & 4,702               & 2,591            & 7,293          \\
D-BR \cite{chen2022dbrst}              & 997                & 656                 & 3,900            & 5,553          \\
EP \cite{zheng2022bnpep}               & 997                & -                   & 53               & 1,050          \\
ReBack \cite{ma2024need}                                 & 997                & 112                 & 92               & 1,201          \\
PIPD \cite{chen2024progressive}                                   & 997                & 1,325               & 438              & 2,760          \\
MSPC \cite{pal2024backdoor}                                   & 997                & 1,844               & 352              & 3,193          \\
\textbf{CGD (ours)}                    & 997                & 60                  & 98               & 1,154          \\
\end{tblr}
\vspace{-0.3cm}
\end{small}
\end{table}

\begin{figure}
\begin{center}
\centerline{\includegraphics[width=0.75\columnwidth]{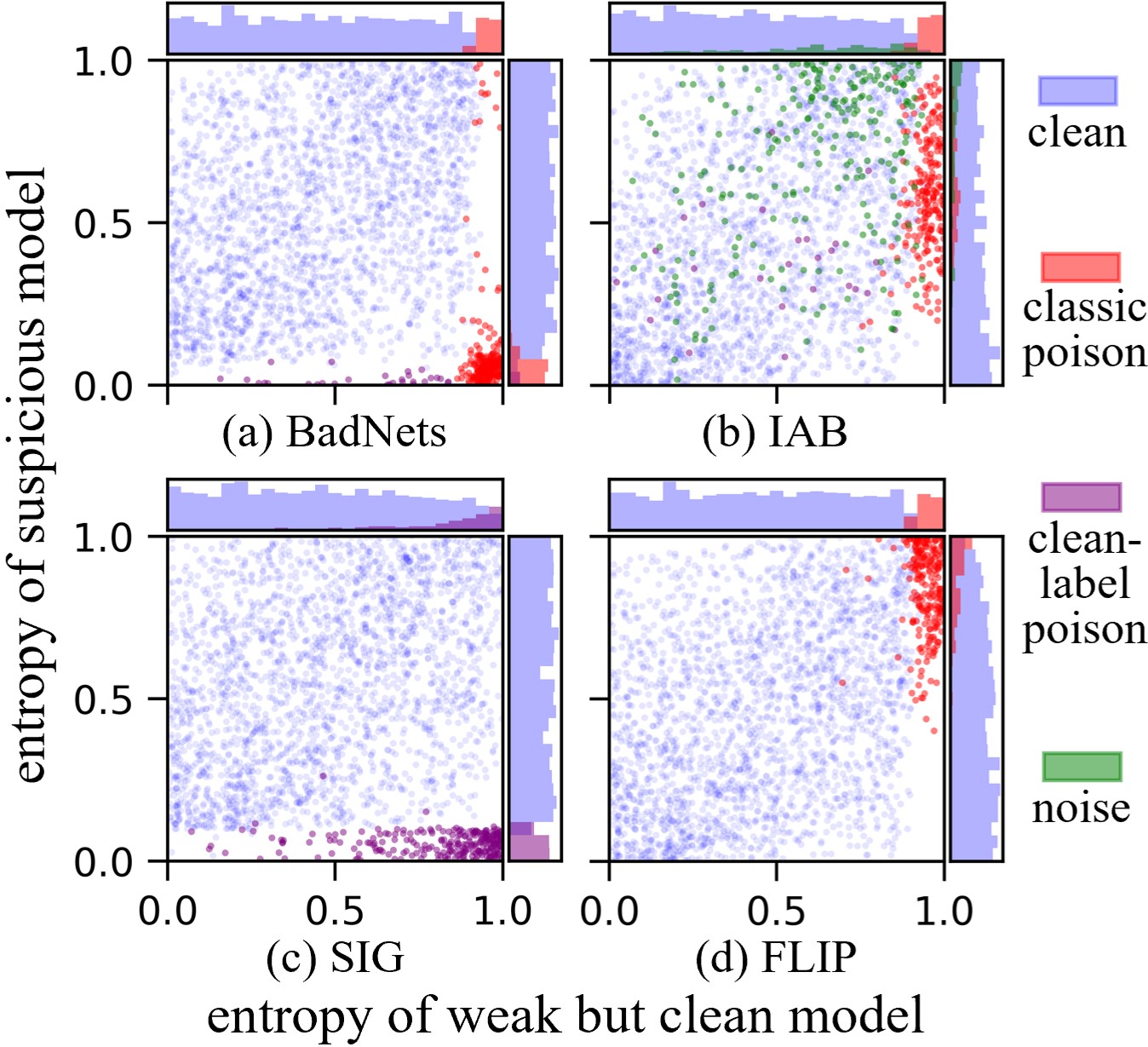}}
\vspace{-0.3cm}
\caption{Entropy distribution plot for various categories of attacks. Poisoned samples (red/purple) are easily separated.}
\label{fig:analysis}
\end{center}
\vspace{-0.7cm}
\end{figure}

\subsection{Main Results}

\subsubsection{\textbf{CLIP-Guided Defense (CGD).}} To evaluate the effectiveness of CGD, we report the CA and ASR results for eight defense methods against eleven attacks on CIFAR-10 in Table \ref{tab:cifar10}. CGD consistently achieves low ASRs ($\leq 1\%$) while maintaining high CAs, with minimal drops ($\leq 0.3\%$) across all defenses. Remarkably, CGD can identify and unlearn poisoned samples, improving CA over the "No Defense" baseline by 3.4\% for WaNet and 3.6\% for IAB. In contrast, self-supervised defenses like DBD, ASD, and D-BR fail to resist clean-label attacks such as CTRL and SIG, while property-based defenses like ABL and EP are vulnerable to clean-image backdoors like GCB and FLIP. Additionally, CGD is highly efficient, requiring less than 3 minutes to split data and apply defenses after training the backdoored model (Table~\ref{tab:time}). While EP \cite{zheng2022bnpep} takes a similar amount of time, it cannot separate clean data. Thus, CGD stands out as the most efficient method for isolating clean data from poisoned samples while maintaining a high defense success rate.

\subsubsection{\textbf{Entropy Map Visualization.}} The entropy map distributions for the four defense categories, shown in Fig. \ref{fig:analysis}
, highlight distinct patterns. (a) \textit{Classical backdoors} (e.g., BadNets \cite{gu2019badnets}): Poisoned samples cluster in the lower right, separable by entropy from suspicious models or CLIP. (b) \textit{Dynamic backdoors} (e.g., IAB \cite{nguyen2020input}): Noise (green points) complicates detection but doesn’t inherently form backdoors. (c) \textit{Clean-label backdoors} (e.g., SIG \cite{barni2019sig}): These resemble classical backdoors in suspicious model entropy, detectable by methods like ABL \cite{li2021anti}. (d) \textit{Clean-image backdoors} (e.g., FLIP \cite{jha2024flip}): High entropy in suspicious models reduces the effectiveness of entropy- or loss-based defenses. Overall, each category’s unique entropy patterns support using category-average performance as a key metric in later experiments.

\begin{table}
\centering
\caption{Clean-data-based defenses. Methods with * use 5\% clean data, while those with the ``-SC" suffix use \textbf{S}plit \textbf{C}lean data extracted from poisoned datasets using CGD.}
\label{tab:clean_data}
\vspace{-0.3cm}
\begin{small}
\begin{tblr}{
  width = 0.85\linewidth,
  rowsep = 0.2pt,
  colsep = 0.2pt,
  colspec = {Q[200]Q[87]Q[87]Q[87]Q[87]Q[87]Q[87]Q[87]Q[87]},
  cells = {c},
  cell{4,6,7,10,11,12,13,14}{9} = {bg = customgreen},
  cell{3,5}{9} = {bg = customred},
  cell{10,12,13,14}{5} = {bg = customgreen},
  cell{3-7,11}{5} = {bg = customred},
  cell{6,10,13}{3} = {bg = customgreen},
  cell{3,4,5,7,11,12,14}{3} = {bg = customred},
  cell{3,5,6,7,11,12}{7} = {bg = customgreen},
  cell{4,10,13,14}{7} = {bg = customred},
  cell{1}{2,4,6,8} = {c=2}{0.174\linewidth},
  cell{8}{2,4,6,8} = {c=2}{0.174\linewidth},
  hline{1,15} = {-}{0.10em},
  hline{2} = {2,4,6,8}{l},
  hline{2} = {3,5,7,9}{r},
  hline{3,8,10} = {-}{0.05em},
  hline{9} = {2,4,6,8}{l},
  hline{9} = {3,5,7,9}{r},
}
\textbf{Defense $\rightarrow$}    & \textbf{FP* \cite{liu2018fp}}    &      & \textbf{MCR* \cite{zhao2020mcr}}    &      & \textbf{I-BAU* \cite{zeng2021ibau}}    &      & \textbf{ANP* \cite{wu2021anp}}    &      \\
\textbf{Attack $\downarrow$}    & CA               & ASR  & CA               & ASR  & CA                 & ASR  & CA              & ASR  \\
Classic        & 92.1             & 21.7 & 92.6             & 51.1 & 86.4               & 16.5 & 85.3               & 21.4 \\
Dynamic        & 92.8             & 30.9 & 72.6             & 47.9 & 90.5               & 30.3 & 87.2               & 1.1  \\
Clean-Label    & 92.4             & 58.0 & 93.3             & 82.5 & 88.7               & 16.0 & 86.5               & 24.9 \\
Clean-Image    & 92.1             & 17.7 & 92.9             & 38.5 & 89.4               & 6.6  & 84.9               & 0.3  \\
Average        & 92.4             & 32.1 & 87.9             & 55.0 & 88.7               & 17.3 & 86.0               & 11.9 \\
\textbf{Defense $\rightarrow$}    & \textbf{FP-SC} &      & \textbf{MCR-SC} &      & \textbf{I-BAU-SC} &      & \textbf{ANP-SC} &      \\
\textbf{Attack $\downarrow$}    & CA               & ASR  & CA               & ASR  & CA                 & ASR  & CA              & ASR  \\
Classic        & 93.3             & 18.2 & 91.0             & 1.4  & 90.7               & 25.7 & 86.6               & 18.8 \\
Dynamic        & 94.0             & 25.2 & 87.6             & 24.3 & 91.8               & 6.2  & 85.7               & 0.6  \\
Clean-Label    & 93.2             & 57.2 & 90.8             & 5.7  & 90.8               & 17.8 & 87.5               & 9.7  \\
Clean-Image    & 92.7             & 15.8 & 91.3             & 1.0  & 90.4               & 43.4 & 85.8               & 0.1  \\
Average        & 93.3             & 29.1 & 90.2             & 8.1  & 90.9               & 23.3 & 86.4               & 7.3  \\
\end{tblr}
\vspace{-0.3cm}
\end{small}
\end{table}

\subsubsection{\textbf{Split Clean-Data Defense.}} We incorporate a portion of selected clean data into clean-data-based defenses and observe a significant improvement in defense performance. Results comparing standard clean-data defenses with their counterparts using CLIP-guided data splitting are presented in Table \ref{tab:clean_data}. As shown, for most methods tested (FP \cite{liu2018fp}, MCR \cite{zhao2020mcr}, ANP \cite{wu2021anp}), our Split Clean (SC) approach achieves superior outcomes, with both lower ASRs and higher CAs. This improvement is likely because our default hyperparameter setting allocates around 70\% of the total data as clean, a significantly larger proportion than the typical clean-data assumption (around 5\% of total data). Consequently, defenses such as ANP and MCR, which are based on the sensitivity of poisoned data, benefit from increased clean data in effectively mitigating backdoor effects.
These results suggest that, for certain defenses, access to only 10\% of poisoned data can yield defense results comparable to using 5\% clean data, underscoring the effectiveness of the Split Clean approach.

\subsubsection{\textbf{Scalability.}} Table \ref{tab:dataset} demonstrates the effectiveness of CGD across three additional datasets: CIFAR-100, GTSRB, and ImageNet. For each dataset, we evaluate all successful backdoor attacks (ASR $\geq$ 90\%) and report the average results across attack categories. In all cases, CGD effectively eliminates backdoors from the models (with ASR reduced to $\leq$ 3\%) while incurring only minimal clean accuracy (CA) reduction (average CA drop $\leq$ 0.3\%). These results confirm the scalability and robustness of CGD across diverse datasets. 

\begin{table}
\centering
\caption{CGD scalability on other datasets.}
\label{tab:dataset}
\vspace{-0.3cm}
\begin{small}
\begin{tblr}{
  width = 0.8\linewidth,
  rowsep = 0.2pt,
  colsep = 0.2pt,
  colspec = {Q[250]Q[113]Q[113]Q[96]Q[119]Q[110]Q[110]},
  cells = {c},
  cell{1}{1} = {r=2}{},
  cell{1}{2} = {c=2}{0.226\linewidth},
  cell{1}{4} = {c=2}{0.215\linewidth},
  cell{1}{6} = {c=2}{0.22\linewidth},
  cell{2}{2} = {c=2}{0.226\linewidth},
  cell{2}{4} = {c=2}{0.215\linewidth},
  cell{2}{6} = {c=2}{0.22\linewidth},
  cell{9}{1} = {r=2}{},
  cell{9}{2} = {c=2}{0.226\linewidth},
  cell{9}{4} = {c=2}{0.215\linewidth},
  cell{9}{6} = {c=2}{0.22\linewidth},
  cell{10}{2} = {c=2}{0.226\linewidth},
  cell{10}{4} = {c=2}{0.215\linewidth},
  cell{10}{6} = {c=2}{0.22\linewidth},
  hline{1,18} = {-}{0.10em},
  hline{3} = {2,4,6}{l},
  hline{3} = {3,5,7}{r},
  hline{11} = {2,4,6}{l},
  hline{11} = {3,5,7}{r},
  hline{4,9,12} = {-}{0.05em},
}
\textbf{Dataset $\rightarrow$} & \textbf{CIFAR-100~}       &      & \textbf{GTSRB~}           &       & \textbf{ImageNet~}        &      \\
                 & \textbf{\textbf{(No Defense)}} &      & \textbf{\textbf{(No Defense)}} &       & \textbf{\textbf{(No Defense)}} &      \\
\textbf{Attack $\downarrow$}  & CA                        & ASR  & CA                        & ASR   & CA                        & ASR  \\
Classic          & 69.6                      & 85.3 & 97.9                      & 96.7  & 56.6                      & 99.1 \\
Dynamic          & 66.2                      & 94.6 & 97.0                      & 95.2  & 57.5                      & 98.7 \\
Clean-Label      & 70.6                      & 38.3 & 98.4                      & 85.5  & 51.6                      & 48.1 \\
Clean-Image      & 67.1                      & 99.8 & 95.3                      & 100.0 & 54.0                      & 99.8 \\
Average          & 68.4                      & 79.5 & 97.1                      & 94.4  & 54.9                      & 86.4 \\
\textbf{Dataset $\rightarrow$} & \textbf{CIFAR-100~}       &      & \textbf{GTSRB~}           &       & \textbf{ImageNet~}        &      \\
                 & \textbf{\textbf{(CGD)}}   &      & \textbf{\textbf{(CGD)}}   &       & \textbf{\textbf{(CGD)}}   &      \\
\textbf{Attack $\downarrow$}  & CA                        & ASR  & CA                        & ASR   & CA                        & ASR  \\
Classic          & 69.6                      & 0.0  & 97.8                      & 0.0   & 56.5                      & 0.0  \\
Dynamic          & 70.4                      & 0.3  & 97.8                      & 1.1   & 57.7                      & 0.0  \\
Clean-Label      & 69.6                      & 2.1  & 94.7                      & 2.8   & 48.7                      & 0.3  \\
Clean-Image      & 68.2                      & 0.0  & 96.8                      & 0.0   & 55.6                      & 0.0  \\
Average          & 69.4                      & 0.6  & 96.8                      & 1.0   & 54.6                      & 0.1  \\ \hline\hline
\textbf{Average Drop}     & \textbf{-1.0}                      &\textbf{78.9}  & \textbf{0.4}                       &\textbf{93.4}  & \textbf{0.3}                       &\textbf{86.4}  
\end{tblr}
\vspace{-0.3cm}
\end{small}
\end{table}

\subsection{Ablation Study}

\subsubsection{\textbf{Loss Term Analysis}}
In our CLIP-guided unlearning approach, we utilize three loss terms: unlearn loss ($\mathcal{L}_{\text{un}}$), relearn loss ($\mathcal{L}_{\text{re}}$), and distillation loss ($\mathcal{L}_{\text{distill}}$). To evaluate their roles, we conducted an ablation study by testing various combinations of these terms, with results shown in Table \ref{tab:abla}, measuring CA and ASR across multiple backdoor categories.

{\ding{182}} \textbf{Single Loss Term.} Using only $\mathcal{L}_{\text{un}}$ yields high CA (average: 88.1) but fails to reduce ASR (average: 93.8), notably against clean-label backdoors (ASR: 98.5), indicating insufficient backdoor mitigation without targeted guidance. Similarly, $\mathcal{L}_{\text{re}}$ alone maintains high CA (average: 92.9) yet struggles with high ASR (average: 86.3), as backdoor knowledge persists. Employing only $\mathcal{L}_{\text{distill}}$ reduces ASR (average: 33.3) but severely degrades CA (average: 34.1), making the model impractical.

{\ding{183}} \textbf{Two-Term Combinations.} Combining $\mathcal{L}_{\text{un}}$ and $\mathcal{L}_{\text{re}}$ improves defense (average ASR: 53.5), yet clean-label backdoors remain potent (ASR: 93.8), highlighting the need for additional guidance. Pairing $\mathcal{L}_{\text{re}}$ and $\mathcal{L}_{\text{distill}}$ reduces ASR significantly (average: 13.5), but dynamic backdoors retain a notable ASR (24.5), showing incomplete protection against sophisticated attacks.

{\ding{184}} \textbf{Synergy of All Three Terms.} Employing all three losses achieves optimal results, with an average CA of 93.1 and an ASR of 0.1 across all backdoor types. Here, $\mathcal{L}{\text{un}}$ targets backdoor unlearning, $\mathcal{L}{\text{distill}}$ aligns the model with correct logits via CLIP’s knowledge, and $\mathcal{L}_{\text{re}}$ preserves clean performance.

\setlength{\intextsep}{5pt} 
\setlength{\columnsep}{5pt}
\begin{wrapfigure}{r}{0.47\linewidth}
\begingroup 
\centering
\vspace{-0.2cm} 
\centerline{\includegraphics[width=\linewidth]{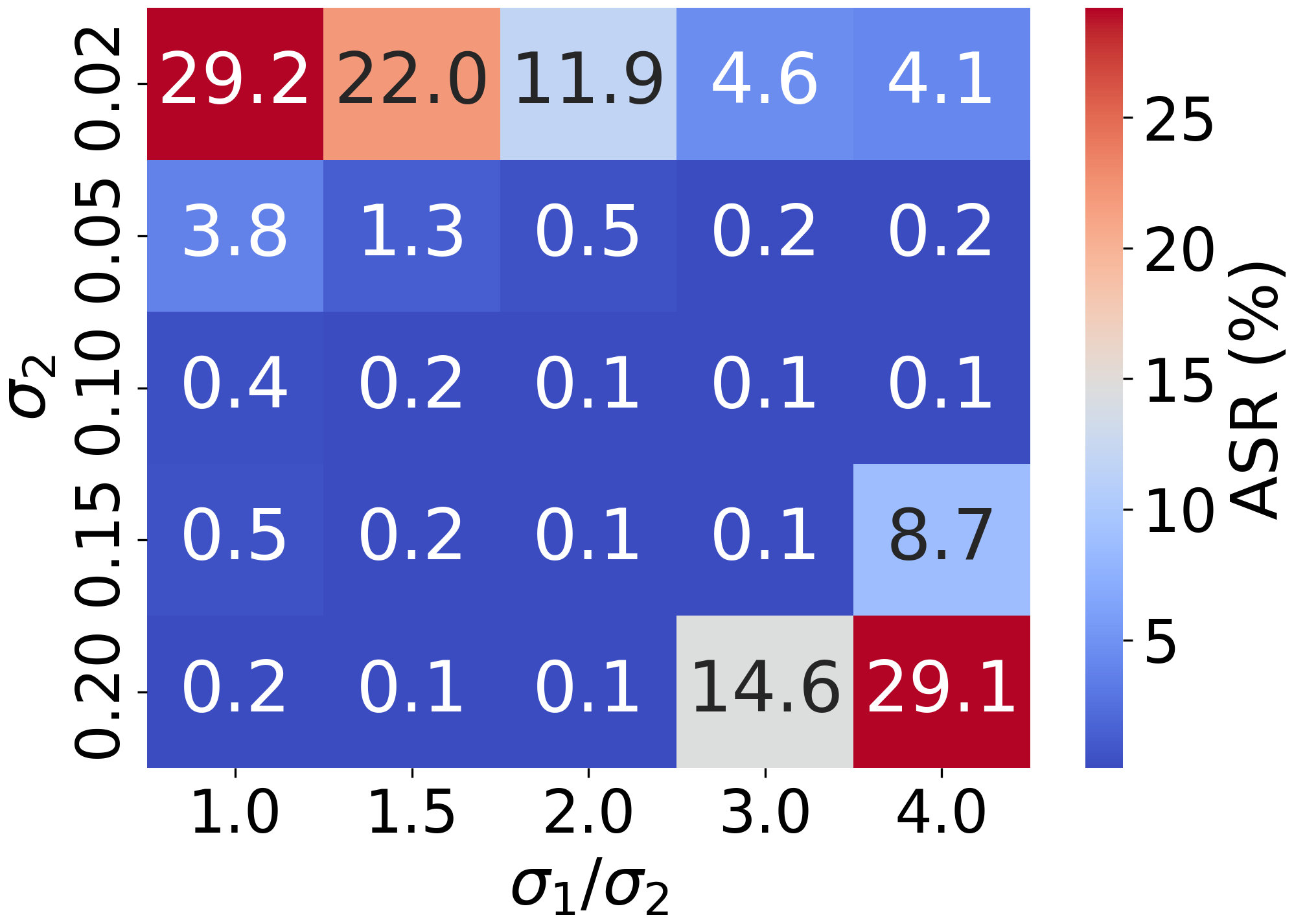}}
\vspace{-0.3cm}
\caption{Threshold Study.}
\vspace{-0.2cm}
\label{fig:sigma}
\endgroup
\end{wrapfigure}

\subsubsection{\textbf{Hyperparameter Study.}} The primary hyperparameters in our experiment are the clean data threshold \(\sigma_1\) and the triggered data threshold \(\sigma_2\), which define the splitting of the entropy map. Selecting suitable values for these thresholds is crucial: if set too low, triggered samples may be misclassified as clean, while overly high settings risk assigning clean samples to the triggered subset. Both cases reduce performance. To maintain \(\sigma_1\) consistently higher than \(\sigma_2\), we optimize using their ratio \(\sigma_1/\sigma_2\). As illustrated in Fig. \ref{fig:sigma}, ASR averaged over 11 attack methods highlights a robust range where ASR remains low. Even in worst-case scenarios, ASR is reduced below 30\%. For default settings, we use \(\sigma_1 = 0.2\) and \(\sigma_2 = 0.1\), though these values can be flexibly adjusted within a reasonable range.

\begin{table}
\centering
\caption{Ablation study for different loss terms in CGD.}
\label{tab:abla}
\vspace{-0.3cm}
\begin{small}
\begin{tblr}{
  width = 1.0\linewidth,
  rowsep = 0.2pt,
  colsep = 0.2pt,
  colspec = {Q[55]Q[55]Q[80]Q[60]Q[60]Q[65]Q[65]Q[74]Q[74]Q[71]Q[71]Q[60]Q[60]},
  cells = {c},
  cell{1}{1} = {c=3}{},
  cell{1}{4} = {c=2}{},
  cell{1}{6} = {c=2}{},
  cell{1}{8} = {c=2}{},
  cell{1}{10} = {c=2}{},
  cell{1}{12} = {c=2}{},
  vline{4,12} = {1-9}{0.05em},
  hline{1,10} = {-}{0.1em},
  hline{3,9} = {-}{0.05em},
  cell{3-4}{5,7,9,11,13} = {bg=customred},
  cell{5}{5} = {bg=customgreen},
  cell{5}{7,9,11,13} = {bg=customred},
  cell{6}{5,7,9,11,13} = {bg=customred},
  cell{7}{5,9,11} = {bg=customgreen},
  cell{7}{7,13} = {bg=customred},
  cell{8}{5,9,11,13} = {bg=customgreen},
  cell{8}{7} = {bg=customred},
  cell{9}{5,7,9,11,13} = {bg=customgreen},
}
Case & & & Classic & & Dynamic & & Clean-label & & Clean-lmage & & Average & \\
$\mathcal{L}_{\text{un}}$ & $\mathcal{L}_{\text{re}}$ & $\mathcal{L}_{\text{distill}}$ & CA & ASR & CA & ASR & CA & ASR & CA & ASR & CA & ASR \\
\checkmark & & & 83.6 & 97.5 & 90.2 & 82.0 & 91.7 & 98.5 & 87.1 & 97.3 & 88.1 & 93.8 \\
 & \checkmark & & \textbf{93.2} & 91.5 & \textbf{93.6} & 62.2 & \textbf{93.6} & 94.4 & 91.2 & 97.2 & 92.9 & 86.3 \\
 & & \checkmark & 26.8 & 13.8 & 51.8 & 64.5 & 34.0 & 29.3 & 24.0 & 25.7 & 34.1 & 33.3 \\
\checkmark & \checkmark & & 93.1 & 48.3 & 93.5 & 34.5 & \textbf{93.6} & 93.8 & 92.0 & 37.2 & 93.0 & 53.5 \\
\checkmark & & \checkmark & 9.3 & 1.1 & 36.8 & 58.9 & 13.5 & 14.3 & 9.7 & 14.3 & 17.3 & 22.1 \\
 & \checkmark & \checkmark & 92.8 & 4.1 & \textbf{93.6} & 24.5 & 92.9 & 11.8 & 92.2 & 13.7 & 92.9 & 13.5 \\
\checkmark & \checkmark & \checkmark & 93.0 & \textbf{0.0} & 93.5 & \textbf{0.3} & 93.4 & \textbf{0.0} & \textbf{92.4} & \textbf{0.2} & \textbf{93.1} & \textbf{0.1} \\
\end{tblr}
\vspace{-0.3cm}
\end{small}
\end{table}

\setlength{\intextsep}{5pt} 
\setlength{\columnsep}{5pt}
\begin{wraptable}{r}{0.46\linewidth}
\centering
\caption{CGD using other language-image pretraining models except CLIP.}
\vspace{-0.35cm}
\label{tab:other_model}
\begin{small}
\begin{tblr}{
  width = 1.0\linewidth,
  rowsep = 0.2pt,
  colsep = 0.2pt,
  colspec = {Q[260]Q[138]Q[138]Q[160]Q[160]},
  cells = {c},
  cell{1}{2} = {c=2}{},
  cell{1}{4} = {c=2}{},
  hline{2} = {2,4}{l},
  hline{2} = {3,5}{r},
  hline{1,8} = {-}{0.1em},
  hline{3,7} = {-}{0.05em},
}
\textbf{Model→}  & \textbf{SigLIP} &     & \textbf{ImageBind} &     \\
\textbf{Attack↓} & CA     & ASR & CA        & ASR \\
Classic & 93.0   & 0.0 & 92.6      & 0.0 \\
Dynamic & 93.2   & 5.3 & 93.3      & 2.5 \\
C-Label & 92.8   & 0.7 & 92.7      & 0.7 \\
C-Image & 90.8   & 3.3 & 90.6      & 2.8 \\
Average & 92.5   & 2.3 & 92.3      & 1.5 
\end{tblr}
\end{small}
\end{wraptable}

\subsubsection{\textbf{Robustness to Different Language-Image Pretraining \\ Models}} Although we mainly use CLIP \cite{radford2021clip} to illustrate CGD's effectiveness, CGD can use other language-image pretraining models to replace CLIP. As shown in Table \ref{tab:other_model}, we consider SigLIP \cite{zhai2023siglip} and ImageBind \cite{girdhar2023imagebind} as two alternative models. The result suggests that both achieve low ASR ($\leq5\%$) while maintaining good CA ($\geq90\%$).

\begin{figure}
  \centering
  \subfloat[CIFAR-100]{
\vspace{-0.15cm}
    \includegraphics[width=0.45\textwidth]{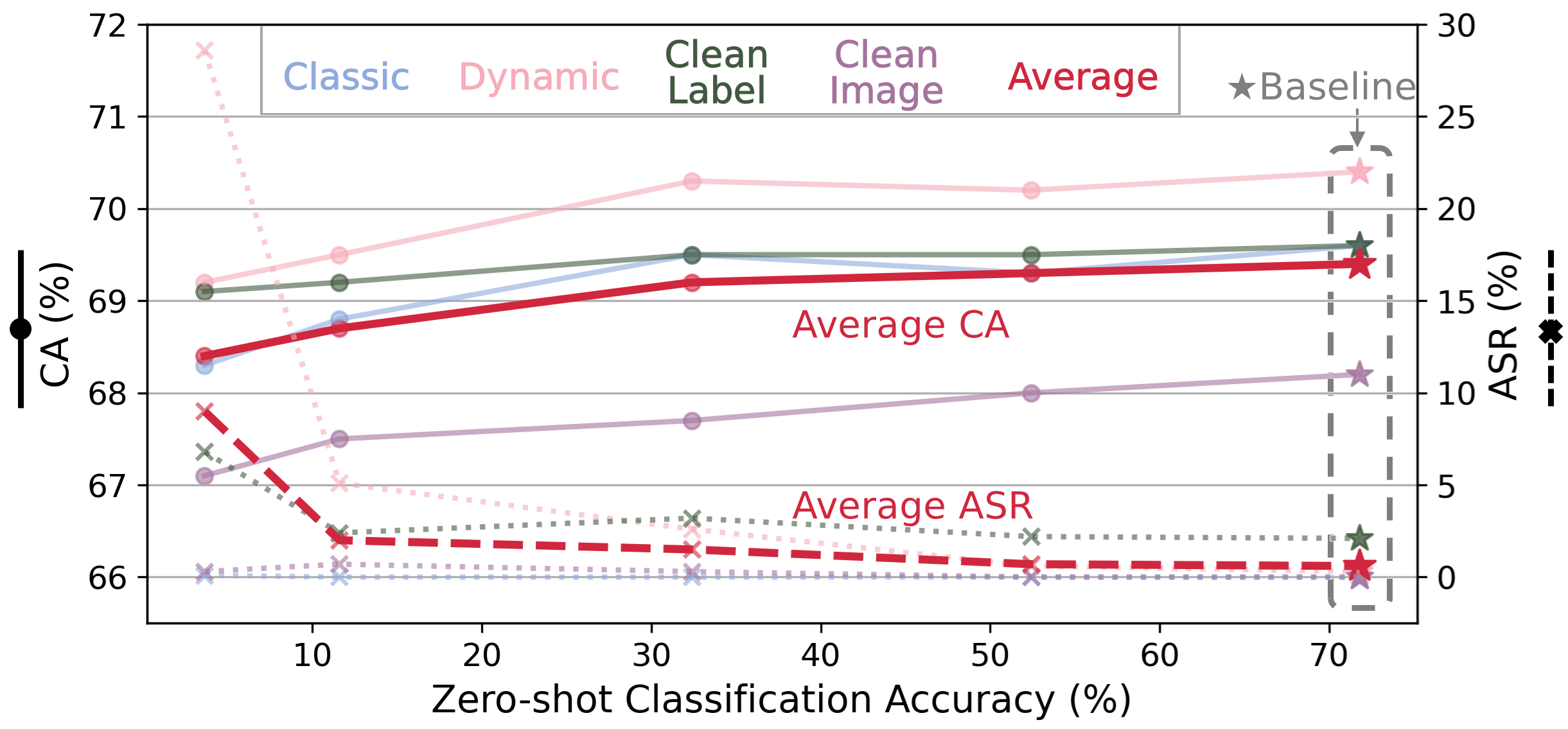}
    \label{fig:weak_cifar}
  }
  \hfill
  \subfloat[TinyImageNet]{
\vspace{-0.15cm}
    \includegraphics[width=0.45\textwidth]{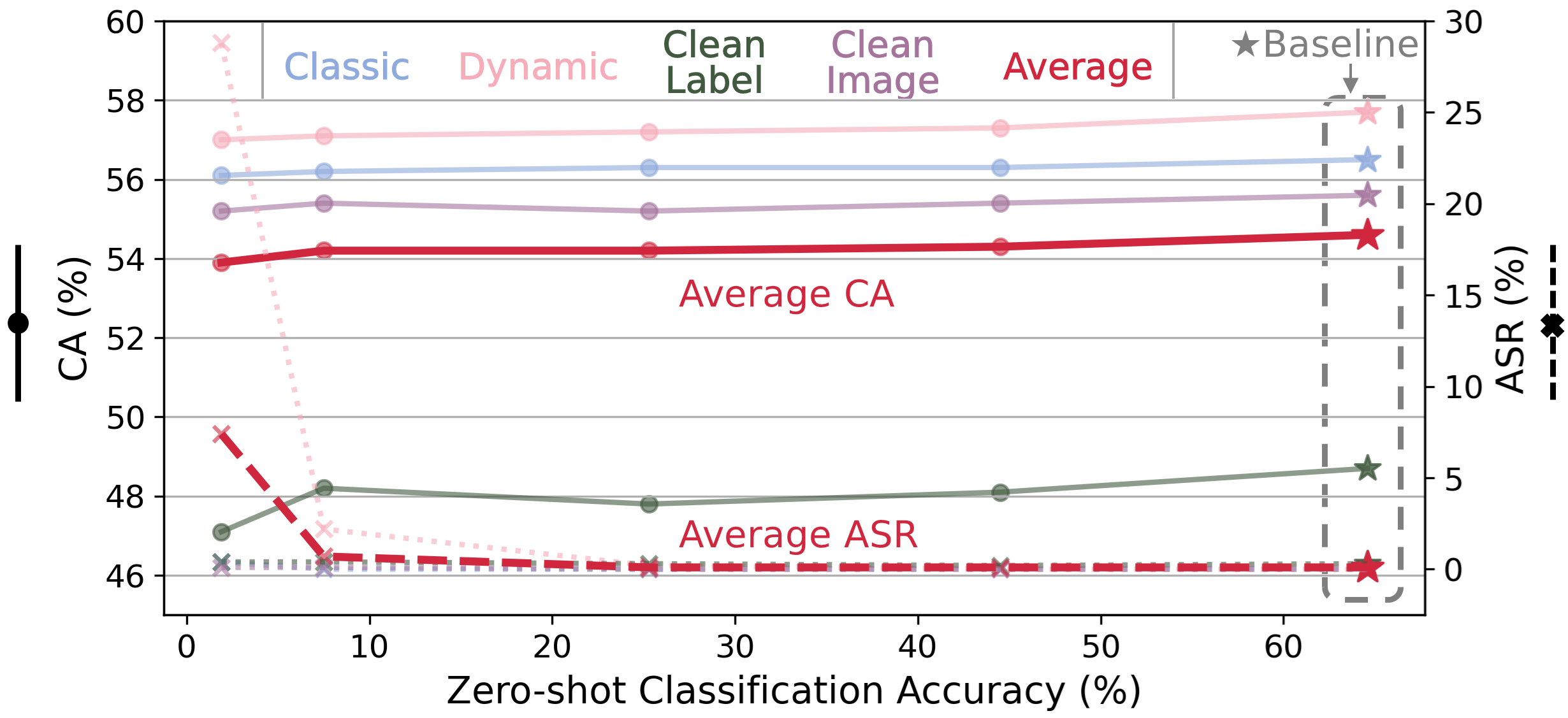}
    \label{fig:weak_tiny}
  }
\vspace{-0.3cm}
  \caption{CGD’s defense performance on low-accuracy CLIP models. CGD successfully defends against all attacks (ASR $\leq 5\%$) even when CLIP's zero-shot accuracy is as low as 10\%.}
  \label{fig:weak}
\vspace{-0.3cm}
\end{figure}

\subsection{Security Guarantee Analysis}

Since our CGD is mainly based on the assumption of CLIP \textit{can} act as an weak but clean model. As a result, there are two major security risks here: (A) \textit{If CLIP is a valid weak model for the specific task}, and (B) \textit{If CLIP is a clean model}. We evaluate these two assumptions one by one.

\subsubsection{\textbf{Weaker CLIP}} We further investigate our defense's robustness when CLIP's accuracy is intentionally reduced. Given that current versions of CLIP are highly effective on common datasets, we manually introduce random noise to CLIP's output, reducing its accuracy in a controlled manner. Specifically, we add Gumbel-distributed noise \cite{jang2016gumbel} to the output logits of CLIP, enabling us to adjust the level of inaccuracy by varying the scale parameter of the Gumbel distribution. As shown in Fig. \ref{fig:weak}, we present results illustrating the defense performance against the zero-shot classification accuracy of CLIP on CIFAR-100 and TinyImageNet. Our CGD approach successfully mitigates all tested attacks (ASR $\leq$ 5\%) when CLIP's zero-shot classification accuracy remains as low as 11.6\% on CIFAR-100 and 7.5\% on TinyImageNet. This is likely because even a weakened CLIP predicts images as similar classes close to the target label, causing the entropy-based calculations from CLIP to become less precise, though still somewhat informative. Additionally, while this noise does not impact backdoor removal capability, it does reduce clean accuracy in tandem with the drop in CLIP accuracy.

To further validate this assumption in real and diverse vision datasets, we carry out very extensive experiments on 17 different datasets, including large-scale datasets like SUN397 and full ImageNet. On these datasets, we tested three representative attacks (Blend, BPP, CTRL) but excluded clean-image backdoors due to either low ASR on large datasets (FLIP) or high computational cost (GCB). As shown in Table \ref{tab:other_dataset}, CGD successfully defends \textbf{all attacks} except on SVHN for BPP. This failure is because CLIP's zero-shot accuracy on SVHN is only 13.4\%, making it thoroughly ineffective for a 10-class prediction dataset. Nonetheless, CGD still defends other SVHN attacks (e.g., Blend, CTRL) via the suspicious model's entropy. Regardless of whether the dataset is large-scale or has low CLIP accuracy (e.g., GTSRB, DTD, and MNIST), CGD remains effective, showing its strong generalizability.

\begin{table}[t]
\centering
\vspace{-0.0cm}
\caption{CGD on 17 datasets against 3 backdoor attacks. ``$\Delta$CA'': change in CA after defenses. CA drops $\leq$4.2\%. CGD fails only when CLIP is thoroughly ineffective (e.g., on SVHN) on dynamic backdoor attack.}
\vspace{-0.2cm}
\label{tab:other_dataset}
\begin{small}
\begin{tblr}{
  width = 1.0\linewidth,
  rowsep = 0.2pt,
  colsep = 0.8pt,
  colspec = {Q[221]Q[92]Q[83]Q[83]Q[83]Q[83]Q[83]Q[83]Q[83]Q[83]},
  cells = {c},
  cell{1}{1} = {r=2}{},
  cell{1}{2} = {r=3}{},  
  cell{1}{3} = {c=2,r=2}{}, 
  cell{1}{5} = {c=6}{},     
  cell{2}{5} = {c=2}{},     
  cell{2}{7} = {c=2}{},     
  cell{2}{9} = {c=2}{},     
  cell{4}{3} = {bg=customblue17},
  cell{4}{4} = {bg=customblue0},
  cell{5}{3} = {bg=customblue16},
  cell{5}{4} = {bg=customblue17},
  cell{6}{3} = {bg=customblue13},
  cell{6}{4} = {bg=customblue0},
  cell{7}{3} = {bg=customblue11},
  cell{7}{4} = {bg=customblue7},
  cell{8}{3} = {bg=customblue11},
  cell{8}{4} = {bg=customblue5},
  cell{9}{3} = {bg=customblue10},
  cell{9}{4} = {bg=customblue0},
  cell{10}{3} = {bg=customblue8},
  cell{10}{4} = {bg=customblue9},
  cell{11}{3} = {bg=customblue8},
  cell{11}{4} = {bg=customblue9},
  cell{12}{3} = {bg=customblue7},
  cell{12}{4} = {bg=customblue7},
  cell{13}{3} = {bg=customblue7},
  cell{13}{4} = {bg=customblue5},
  cell{14}{3} = {bg=customblue7},
  cell{14}{4} = {bg=customblue5},
  cell{15}{3} = {bg=customblue6},
  cell{15}{4} = {bg=customblue2},
  cell{16}{3} = {bg=customblue3},
  cell{16}{4} = {bg=customblue1},
  cell{17}{3} = {bg=customblue2},
  cell{17}{4} = {bg=customblue1},
  cell{18}{3} = {bg=customblue2},
  cell{18}{4} = {bg=customblue5},
  cell{19}{3} = {bg=customblue2},
  cell{19}{4} = {bg=customblue1},
  cell{20}{3} = {bg=customblue0},
  cell{20}{4} = {bg=customblue0},
  cell{4-20}{6} = {bg = customgreen},
  cell{5-20}{8} = {bg = customgreen},
  cell{4}{8} = {bg = customred},
  cell{4-20}{10} = {bg = customgreen},
  hline{3} = {3,5,7,9,11}{l},
  hline{3} = {4,6,8,10,12}{r},
  hline{2} = {5-11}{},
  hline{1,21} = {-}{0.1em},
  hline{4} = {-}{0.05em},
}
\textbf{Task→}         & {\textbf{Data}\\\textbf{Size}} & {\textbf{Classify}\\\textbf{Accuracy}} &        & \textbf{CGD (Ours)} &       &        &        &             &       \\
              &                &                &        & \textbf{Blend}              &       & \textbf{BPP} &        & \textbf{CTRL} &       \\
\textbf{Dataset↓}      &               & CLIP           & RN50   & $\Delta$CA                  & ASR   & $\Delta$CA    & ASR    & $\Delta$CA         & ASR   \\
SVHN          & 234M         & 13.4         & 95.7 & 0.5                & 0.1 & -1.4 & 100  & -4.2      & 8.6 \\
Country211    & 20.9G        & 17.2         & 12.2 & 0.9                & 0.1 & 0.6  & 0.0  & 0.4       & 9.7 \\
GTSRB         & 689M         & 32.6         & 98.5 & -1.0               & 0.0 & 2.2  & 0.7  & -3.8      & 0.2 \\
FER2013       & 950M         & 41.4         & 62.1 & -0.4               & 0.0 & 3.1  & 0.1  & -0.4      & 0.0 \\
DTD           & 1.2G         & 44.3         & 71.8 & 1.2                & 0.0 & 3.2  & 1.1  & -2.2      & 0.7 \\
MNIST         & 127M         & 48.2         & 99.6 & -2.3               & 7.8 & 0.2  & 0.0  & 0.0       & 0.0 \\
RenderedSST2  & 305M         & 58.6         & 53.8 & 2.8                & 0.0 & 4.8  & 0.5  & -1.8      & 0.0 \\
StanfordCars  & 3.7G         & 59.6         & 52.7 & 0.6                & 0.0 & 7.0 & 0.0  & 2.8       & 0.5 \\
SUN397        & 73.6G        & 62.5         & 61.5 & -0.4               & 0.0 & 2.0 & 0.0  & 0.2       & 5.5 \\
ImageNet      & 146G         & 63.3         & 74.3 & -0.3               & 0.0 & 0.1  & 0.0  & -3.1      & 0.5 \\
CIFAR100      & 338M         & 64.2         & 73.6 & -0.4               & 0.0 & 4.9  & 0.1  & -2.8      & 5.8 \\
Flowers102    & 676M         & 66.5         & 89.5 & -1.3               & 0.2 & 4.2  & 0.8  & -1.3      & 0.4 \\
Caltech101    & 324M         & 81.6         & 92.0 & -2.9               & 3.6 & 5.5  & 3.0  & -3.3      & 0.0 \\
OxfordIIITPet & 1.6G         & 87.3         & 92.9 & 0.8                & 0.3 & 8.3 & 12.4 & 1.9       & 0.0 \\
Food101       & 5.3G         & 88.8         & 72.6 & 7.2                & 0.0 & 2.4 & 0.0  & 0.1       & 8.3 \\
CIFAR10       & 340M         & 89.8         & 95.4 & -0.5               & 0.0 & 2.6  & 0.3  & 0.0       & 0.1 \\
STL10         & 5.4G         & 97.1         & 97.5 & 3.0                & 0.0 & 4.9  & 0.2  & 2.1       & 0.0 
\end{tblr}
\end{small}
\vspace{-0.2cm}
\end{table}

\setlength{\intextsep}{5pt} 
\setlength{\columnsep}{10pt}
\begin{wraptable}{r}{0.4\linewidth}
\begingroup 
\setlength{\tabcolsep}{4pt} 
\centering
\caption{CGD against backdoored CLIPs.}
\vspace{-0.35cm}
\label{tab:backdoorclip}
\begin{small}
\begin{tabular}{c c c c}
\specialrule{0.1em}{0pt}{0pt} 
 & \textbf{(i)} & \textbf{(ii)} & \textbf{(iii)} \\
\specialrule{0.05em}{0pt}{0pt} 
\textbf{CA}     & 92.8 & 92.3 & 92.3 \\
\textbf{ASR-O}  & 0.0  & 0.0  & 0.1  \\
\textbf{ASR-C}  & 0.0  & 0.6  & 0.1  \\
\specialrule{0.1em}{0pt}{0pt} 
\end{tabular}
\end{small}
\endgroup
\end{wraptable}

\subsubsection{\textbf{Backdoored CLIP Analysis.}} Previous research has shown that CLIP models can be vulnerable to backdoor attacks \cite{carlini2022badclip}. In this study, we investigate whether a backdoored CLIP model could pass its backdoor to a victim model during our defense process. We explore three cases: (i) \textit{CLIP has a different backdoor trigger}; (ii) \textit{CLIP has the same trigger but targets a different class}; and (iii) \textit{CLIP has the same trigger and targets the same class}. For each case, we use the BadNets trigger \cite{gu2019badnets} (aligned with \cite{carlini2022badclip}) and measure two metrics: ASR-O (attack success rate for the suspicious dataset’s trigger) and ASR-C (attack success rate for CLIP’s trigger).

Our findings, shown in Table \ref{tab:backdoorclip}, reveal that even when CLIP is backdoored, it successfully removes backdoors from the victim model without transferring its own backdoor. This is mainly because the BadNets trigger, a simple and fixed pattern, can be detected by analyzing entropy in either CLIP or the victim model. As a result, the backdoor in CLIP does not compromise the defense, as the victim model’s entropy effectively identifies the trigger. However, in cases (ii) and (iii), we notice a small decrease in CA. This likely occurs because the poisoned outputs from CLIP introduce slight errors during the guidance process, mildly affecting CA.

\subsection{Resistance to Potential Adaptive Attacks}

\subsubsection{\textbf{Threat Model.}} In the previous experiments, we assumed that attackers have no knowledge of our backdoor defense. In this section, we explore a more challenging scenario where attackers are aware of our defense strategy and adjust their poisoning strategies accordingly. However, they still cannot control the training process after injecting poisoned samples.

\subsubsection{\textbf{Methods.}} Our CGD examines the poisoned dataset using both the suspicious model and a CLIP model. Attackers may evade detection by: (1) creating clean-label backdoors with high entropy in the suspicious model or (2) creating dynamic backdoors with low entropy in the CLIP model.
To counter (1), we use an adaptive SIG-based trigger, \textit{Ada-SIG} \cite{qi2022revisiting}, enhanced by two techniques: Random Patch Masking (\textit{Rand-Patch}), which divides the trigger into 16 patches and randomly masks 50\%, and Random Opacity (\textit{Rand-Opacity}), setting the trigger opacity to 50\%. These methods make the trigger harder to detect. 
For (2), we apply the \textit{Feature Mixing Backdoor} (FMB) attack \cite{lin2020composite}, which mixes features from two classes to target a third class, confusing the CLIP model with multiple class features in one image.

\subsubsection{\textbf{Results.}} Table \ref{tab:adaptive} demonstrates that our CGD defense effectively mitigates both adaptive attacks. Against Adaptive-SIG, CGD reduces ASRs to $\leq 2.4\%$. As shown in Fig. \ref{fig:adasig}, while some triggered samples evade detection during the splitting stage, they retain minimal trigger information, with strongly triggered samples all identified and unlearned. For the FMB attack, ASR is also reduced to $\leq 1.2\%$ across both one-to-one and all-to-one settings. This occurs because, although the CLIP model may confuse mixed-feature samples, it only misclassifies them within the two mixed classes, not the target (third) class. Consequently, the poisoned target label also yields a high cross-entropy score, making the attack ineffective against our defense. 

\begin{table}
\centering
\caption{Potential adaptive attacks against our CGD. }
\label{tab:adaptive}
\vspace{-0.2cm}
\begin{small}
\begin{tblr}{
  width = 0.8\linewidth,
  rowsep = 0.8pt,
  colsep = 0.8pt,
  colspec = {Q[250]Q[250]Q[110]Q[110]Q[110]Q[110]},
  cells = {c},
  cell{1}{1} = {c=2}{},
  cell{1}{3} = {c=2}{},
  cell{1}{5} = {c=2}{},
  cell{2}{1} = {c=2}{},
  cell{3}{1} = {r=3}{},
  cell{6}{1} = {r=2}{},
  hline{1,8} = {-}{0.1em},
  hline{2} = {3,5}{l},
  hline{2} = {4,6}{r},
  hline{3,6} = {-}{0.05em},
}
\textbf{Defense →}  &   & \textbf{No Defense} &         & \textbf{CGD (Ours)} &       \\
\textbf{Adaptive Attacks ↓}     & & CA        & ASR     & CA         & ASR   \\
Ada-SIG          & Rand Patch   & 93.5     & 100.0 & 93.5     & 1.0 \\
                      & Rand Opacity & 93.8     & 91.1  & 93.1     & 1.7 \\
                      & Both Rand    & 93.7     & 95.9  & 93.2     & 2.4 \\
FMB                   & One-to-One   & 92.0     & 85.0  & 92.0     & 0.3 \\
                      & All-to-One   & 91.5     & 77.7  & 91.2     & 1.2 
\end{tblr}
\end{small}
\vspace{-0.2cm}
\end{table}

\section{Conclusion}

We introduce CGD, a novel poison-data-based backdoor defense that leverages CLIP’s zero-shot capabilities to detect and mitigate attacks. Using an entropy-based method to separate poisoned data and guide model retraining, CGD reduces attack success rates below 1\% across various datasets and attack types while maintaining high clean accuracy. Extensive experiments show CGD’s superior efficiency and resilience, even with weaker or backdoored CLIP models. CGD outperforms existing defenses and enhances clean-data-based methods, providing a practical and scalable solution for securing real-world deep learning applications.


\bibliographystyle{ACM-Reference-Format}
\bibliography{main}

\appendix

\appendix


\section{Conditional Analysis of CLIP-Guided Backdoor Detection}

\label{sec:math}

This appendix provides an idealized finite-sample guarantee for the
poisoned-data splitting rule of CGD under independent scoring.  The guarantee
is deliberately conditional: model calibration alone does not imply that
clean and poisoned examples have separable score distributions.  Instead, we
state the score-separation conditions required by CGD and show that, when the
scoring functions are fixed independently of the evaluated examples, the
empirical detection rates concentrate around their population values.

\subsection{Setup and Score-Separation Conditions}

Let \(C\), \(P_L\), and \(P_{CL}\) denote the clean, label-poisoned, and
clean-label-poisoned subsets, respectively.  CGD computes the label-conditioned
negative log-likelihood scores
\begin{align*}
\mathcal{S}^{\mathrm{CLIP}}(x_i)
&=-\log p_{\mathrm{CLIP}}(y_i\mid x_i),\\
\mathcal{S}^{\mathrm{Model}}(x_i)
&=-\log p_{\mathrm{Model}}(y_i\mid x_i).
\end{align*}
Throughout Theorem~1, \(p_{\mathrm{CLIP}}\) and
\(p_{\mathrm{Model}}\), and hence both score functions, are treated as fixed
independently of the examples on which recall and false-positive rates are
evaluated.  This condition is satisfied, for example, when the scoring model
is trained on a separate data split and evaluated on an independent audit
split.  It is not automatically satisfied when a model is trained and scored
on the same examples.

Let \(R^{\mathrm{CLIP}},R^{\mathrm{Model}}\in[0,1]\) be their population
percentile ranks under the data-mixture distribution.  For a triggered-data
threshold \(\sigma_2\), define
\[
A(x)=\mathbf{1}\!\left[R^{\mathrm{CLIP}}(x)>1-\sigma_2\right],
\qquad
B(x)=\mathbf{1}\!\left[R^{\mathrm{Model}}(x)<\sigma_2\right].
\]
The population counterpart of the CGD suspicious-set rule is
\(Z(x)=A(x)\lor B(x)\).

We make the following explicit score-separation assumptions:
\begin{align}
\mathbb{P}(A=1\mid P_L)&\geq 1-\alpha_L,
&\mathbb{P}(B=1\mid P_{CL})&\geq 1-\alpha_{CL}, \label{eq:poison-separation}\\
\mathbb{P}(A=1\mid C)&\leq \beta_A,
&\mathbb{P}(B=1\mid C)&\leq \beta_B. \label{eq:clean-separation}
\end{align}
Here, \(\alpha_L\) and \(\alpha_{CL}\) are miss-rate parameters, while
\(\beta_A\) and \(\beta_B\) control clean false positives.  These quantities
describe ranking quality and are not implied by Expected Calibration Error.
They are population-level parameters and are not known to the defender.  In a
controlled evaluation, they may be estimated using an independently labeled
audit set together with finite-sample confidence intervals.  No independence
between \(A\) and \(B\) is required.

For target-class clean-label poisoned training examples, \(y_i=t\).  When a
trigger increases the victim model's confidence in the target class,
\(p_{\mathrm{Model}}(y_i\mid x_i)\) increases and hence
\(-\log p_{\mathrm{Model}}(y_i\mid x_i)\) decreases.  This motivates the
second condition in Eq.~\eqref{eq:poison-separation}.  For clean-label variants
that do not produce this low-score behavior, the condition need not hold and
the guarantee below does not claim successful detection.

\subsection{Finite-Sample Detection Guarantee}

\noindent\textbf{Theorem 1 (conditional detection guarantee).}
Suppose examples are independent within each of \(C\), \(P_L\), and
\(P_{CL}\).  Suppose further that both scoring functions, their population
percentile functions, and the thresholds are fixed independently of the
evaluated examples.  Let the three sample sizes be \(n_C\), \(n_L\), and
\(n_{CL}\).  Under Eqs.~\eqref{eq:poison-separation}--\eqref{eq:clean-separation},
for every \(\varepsilon>0\),
\begin{align}
\mathbb{P}\!\left(
\widehat{\operatorname{Recall}}_L<1-\alpha_L-\varepsilon
\right)
&\leq \exp(-2n_L\varepsilon^2), \label{eq:label-recall-bound}\\
\mathbb{P}\!\left(
\widehat{\operatorname{Recall}}_{CL}<1-\alpha_{CL}-\varepsilon
\right)
&\leq \exp(-2n_{CL}\varepsilon^2), \label{eq:cl-recall-bound}\\
\mathbb{P}\!\left(
\widehat{\operatorname{FPR}}>\beta_A+\beta_B+\varepsilon
\right)
&\leq \exp(-2n_C\varepsilon^2). \label{eq:fpr-bound}
\end{align}

\noindent\textit{Proof.}
For a label-poisoned example, \(Z=A\lor B\geq A\), and therefore
\(\mathbb{E}[Z\mid P_L]\geq1-\alpha_L\).  Hoeffding's inequality for
independent bounded variables gives Eq.~\eqref{eq:label-recall-bound}.  The same
argument using \(Z\geq B\) gives Eq.~\eqref{eq:cl-recall-bound}.  For clean
examples, the union bound gives
\[
\mathbb{P}(Z=1\mid C)
\leq\mathbb{P}(A=1\mid C)+\mathbb{P}(B=1\mid C)
\leq\beta_A+\beta_B.
\]
A final application of Hoeffding's inequality gives Eq.~\eqref{eq:fpr-bound}.
\hfill\(\square\)

The theorem analyzes the same union operation used by CGD and does not require
knowledge of the poisoning rate.  However, it applies exactly only in the
independent-scoring setting stated above.  It guarantees only the quality of
the data split under the score-separation conditions; it does not guarantee
the final attack success rate or clean accuracy after fine-tuning.

\subsection{Empirical Percentile Ranks under Independent Scoring}

Consider an independent-scoring variant that uses empirical rather than
population percentile ranks.  Let \(\widehat F\) be the empirical CDF of either
fixed score function and \(F\) its population CDF.  Under i.i.d. mixture
sampling, the Dvoretzky--Kiefer--Wolfowitz inequality gives
\[
\mathbb{P}\!\left(\sup_s|\widehat F(s)-F(s)|>\delta\right)
\leq 2\exp(-2N\delta^2).
\]
Applying a union bound to the two score functions, both empirical percentile
ranks are uniformly within \(\delta\) of their population counterparts with
probability at least \(1-4\exp(-2N\delta^2)\).  Consequently, for
\(0<\delta<\sigma_2\), Theorem~1 also applies to empirical ranks after
introducing a \(\delta\)-margin: the poison separation conditions are evaluated
at the stricter thresholds
\(1-\sigma_2+\delta\) and \(\sigma_2-\delta\), while the clean false-positive
conditions are evaluated at the relaxed thresholds
\(1-\sigma_2-\delta\) and \(\sigma_2+\delta\).  The DKW failure probability is
then added to the right-hand side of each corresponding bound above.

In the same-data implementation used in our experiments, the intermediate
model is trained and scored on the suspicious training set.  Its scores may
therefore be statistically dependent across examples, so the independence
condition required by Theorem~1 and the DKW extension is not guaranteed.  We
interpret the theorem as an idealized conditional analysis of the splitting
rule and support the same-data implementation empirically.  Establishing a
finite-sample guarantee for that implementation would additionally require an
algorithmic-stability or dependent-data analysis.  Coordinated or adaptive
poisoning may also violate the score-separation assumptions and is evaluated
separately; data-free online auditing offers a complementary setting~\cite{xu2026internal}.
Stateful dependence also arises in LLM agents that self-manage context~\cite{xu2026llm}.

\section{Detailed Analysis of Resistance to Adaptive Attacks}

\label{ap:adaptive}

In our main experiments, we operated under the assumption that attackers are unaware of our backdoor defense mechanisms. Here, we explore a more challenging scenario in which attackers have complete knowledge of our defense strategies and adapt their poisoning techniques accordingly.

\subsection{Threat Model and Adaptive Attack Strategies}

\subsubsection{\textbf{Threat Model.}} Consistent with prior research \cite{chen2017blended, gu2019badnets, turner2019label}, we assume attackers can access the entire training dataset and are informed about the defense mechanisms we employ. However, they lack control over the training process beyond injecting poisoned samples. Trace-economic underwriting offers a complementary view of autonomous-agent deployment risk~\cite{xu2026agent}.

\subsubsection{\textbf{Adaptive Attack Strategies.}} Our defense relies on examining both the suspicious model and a CLIP model, providing two potential avenues for attackers to attempt evasion:

\begin{enumerate}
    \item \textbf{Enhancing Clean-Label Backdoors' Entropy in the Suspicious Model.} Attackers aim to design triggers that produce higher entropy outputs in the suspicious model, making them less detectable. We incorporate the concept of Adaptive-Blend \cite{qi2022revisiting} with the SIG trigger \cite{barni2019sig}, employing two enhancement techniques:
    \begin{itemize}
        \item \textbf{Random Patch Masking (Rand-Patch):} The trigger is divided into 16 patches, with 50\% randomly masked during training. This randomization reduces the trigger's visibility and consistency.
        \item \textbf{Reduced Opacity (Rand-Opacity):} The trigger's opacity is set to 50\%, making it less prominent in poisoned images.
    \end{itemize}
    These methods aim to create stealthier triggers that are harder for our defense to isolate from benign samples.

    \item \textbf{Reducing Dynamic Backdoors' Entropy in the CLIP Model.} Attackers attempt to lower the cross-entropy scores in the CLIP model by using the Feature Mixing Backdoor (FMB) attack \cite{lin2020composite}. This attack mixes features from different classes to create poisoned samples:
    \begin{itemize}
        \item For instance, combining features from classes A and B to misclassify them as class C, while ensuring that unmixed features (e.g., A+A) are correctly classified.
    \end{itemize}
    The FMB attack introduces ambiguity in the CLIP model's predictions, potentially evading detection based on high cross-entropy scores.
\end{enumerate}

Both adaptive attacks utilize a poison rate of 5\%. For the FMB attack, an additional 50\% of the samples are used for self-mixing noise to enhance stealthiness.

\begin{figure}
\begin{center}
\centerline{\includegraphics[width=\columnwidth]{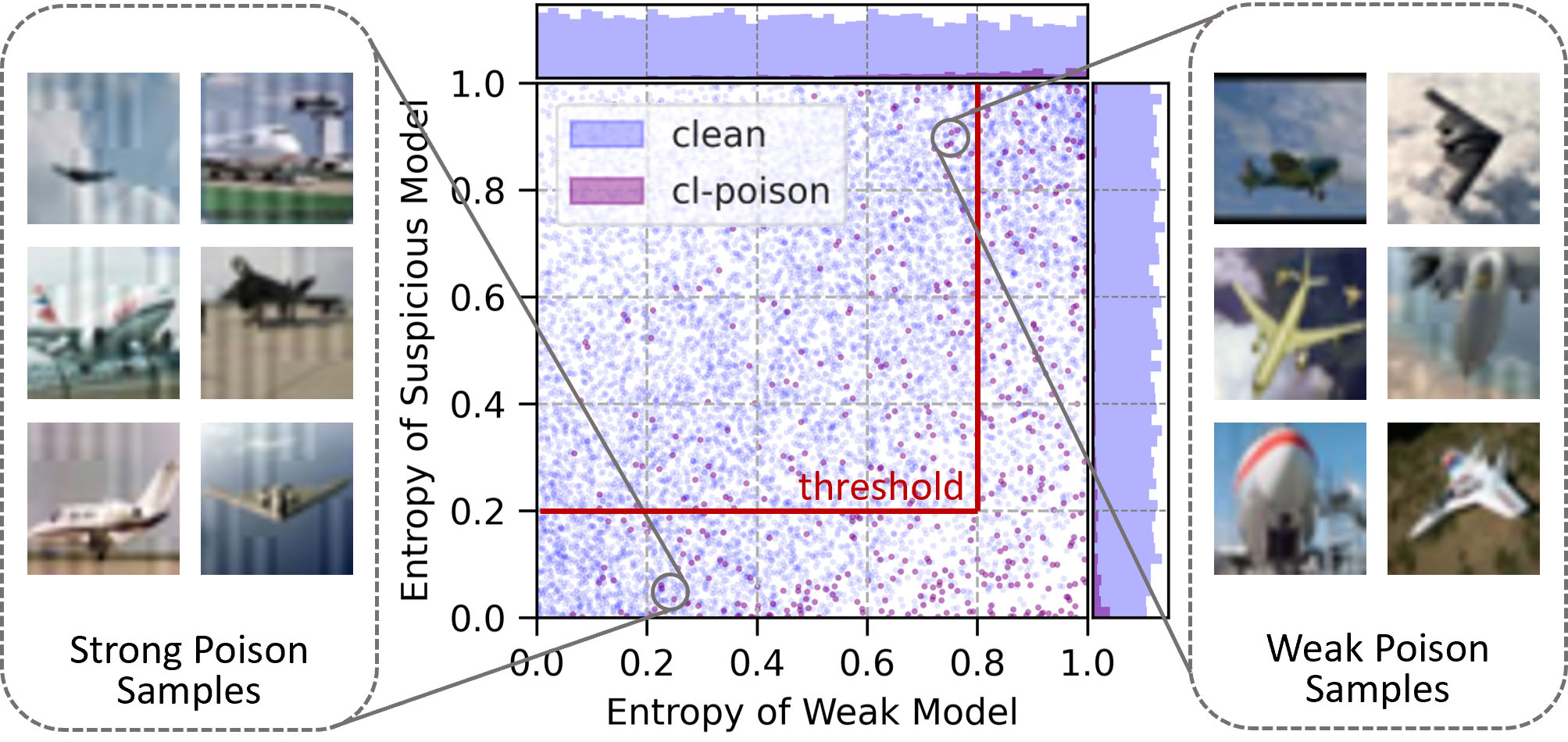}}
\caption{Entropy map for Adaptive SIG with both random patch and random opacity. Although many samples fall outside the detection region, they are weakly poisoned samples with low trigger strength, while most strongly triggered samples are captured in the evaluated runs.}
\label{fig:adasig}
\end{center}
\vspace{-0.3cm}
\end{figure}

\begin{figure}
\begin{center}
\centerline{\includegraphics[width=\columnwidth]{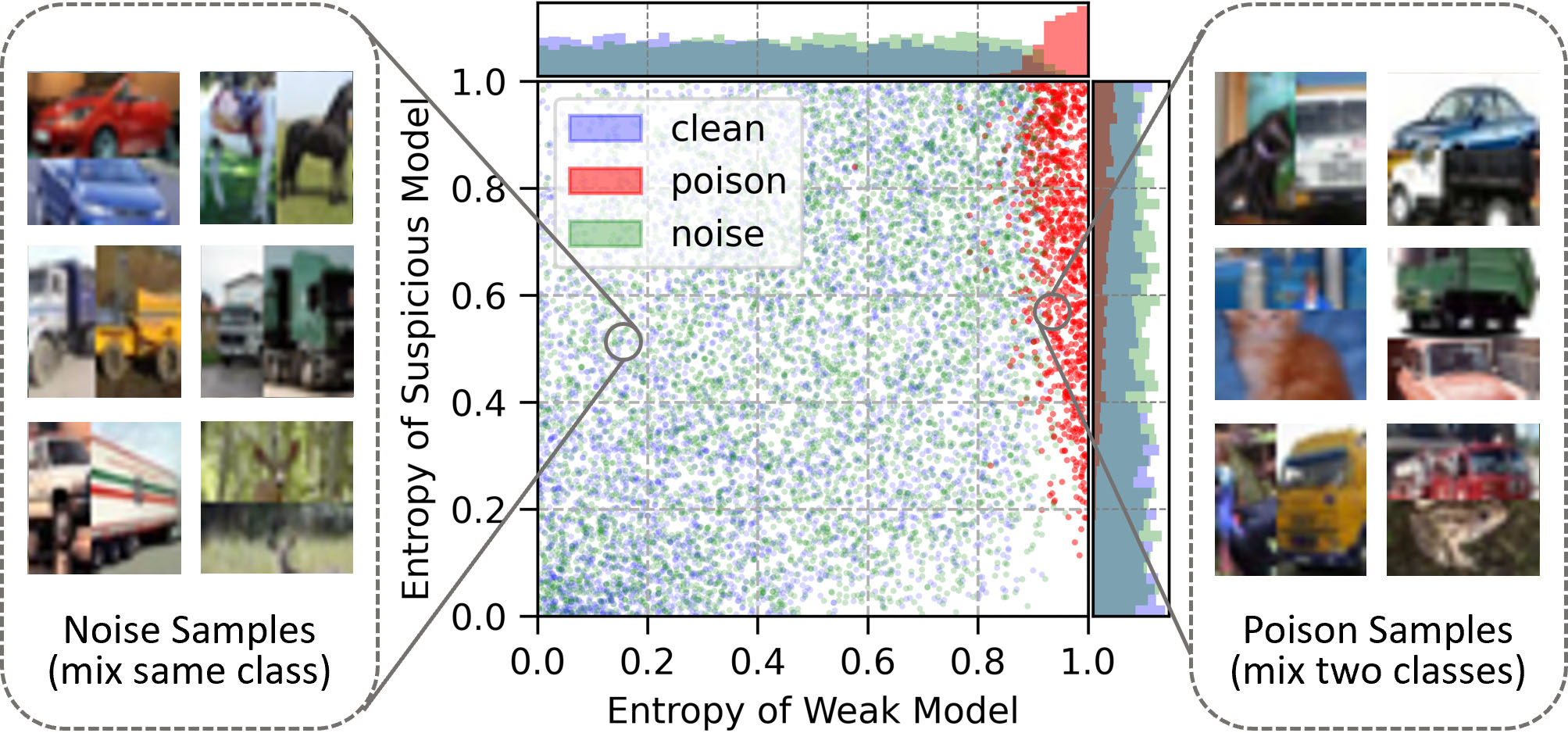}}
\caption{Entropy map for Feature Mixing Backdoor (FMB) under the all-to-one setting: Data from the same class is considered its original label, while data from different classes is assigned the target label. Despite the combined features, all poisoned samples are easily identifiable under CLIP examination, leading to the success of our CGD.}
\label{fig:fmb}
\end{center}
\vspace{-0.3cm}
\end{figure}

\begin{figure*}[ht]
\vskip 0.0in
\centering

\newlength{\subfigwidth}
\setlength{\subfigwidth}{0.21\linewidth} 

\begin{subfigure}[b]{\subfigwidth}
    \centering
    \includegraphics[width=\linewidth]{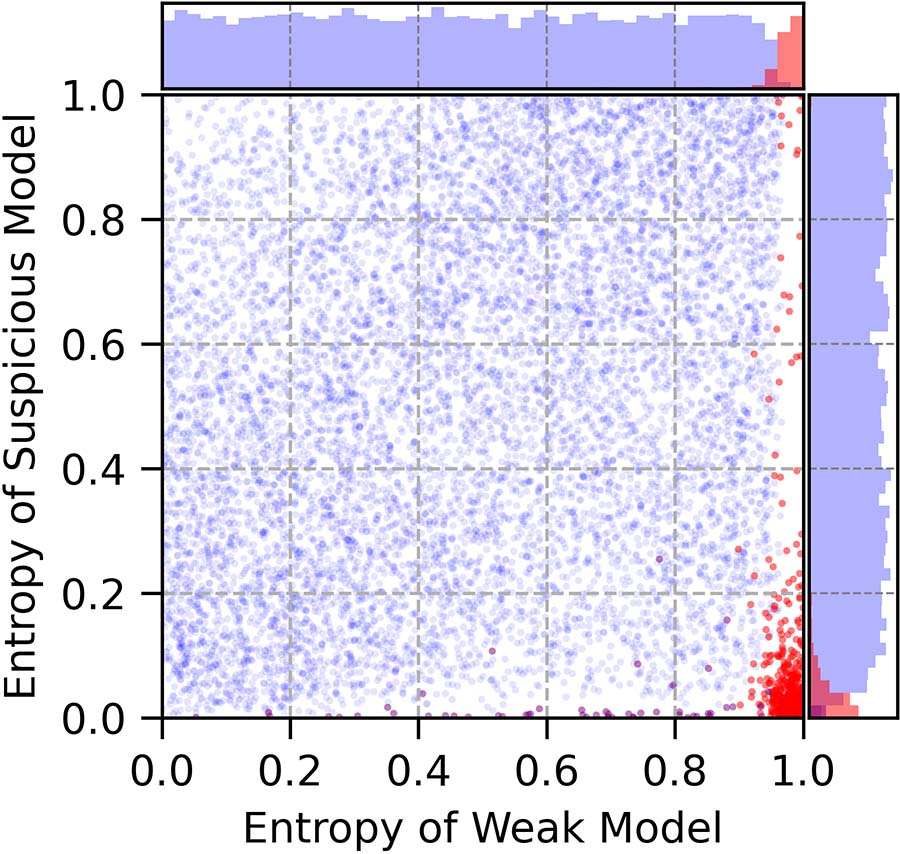}
    \caption{BadNets}
    \label{fig:badnet}
\end{subfigure}
\begin{subfigure}[b]{\subfigwidth}
    \centering
    \includegraphics[width=\linewidth]{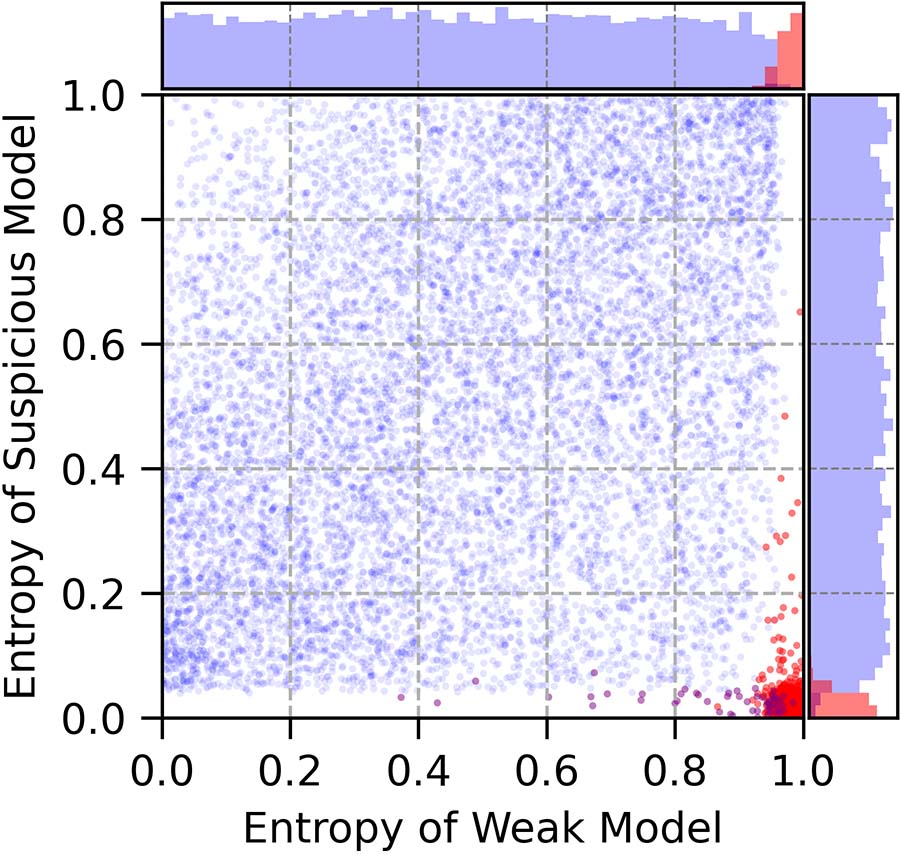}
    \caption{Blend}
    \label{fig:blended}
\end{subfigure}
\begin{subfigure}[b]{\subfigwidth}
    \centering
    \includegraphics[width=\linewidth]{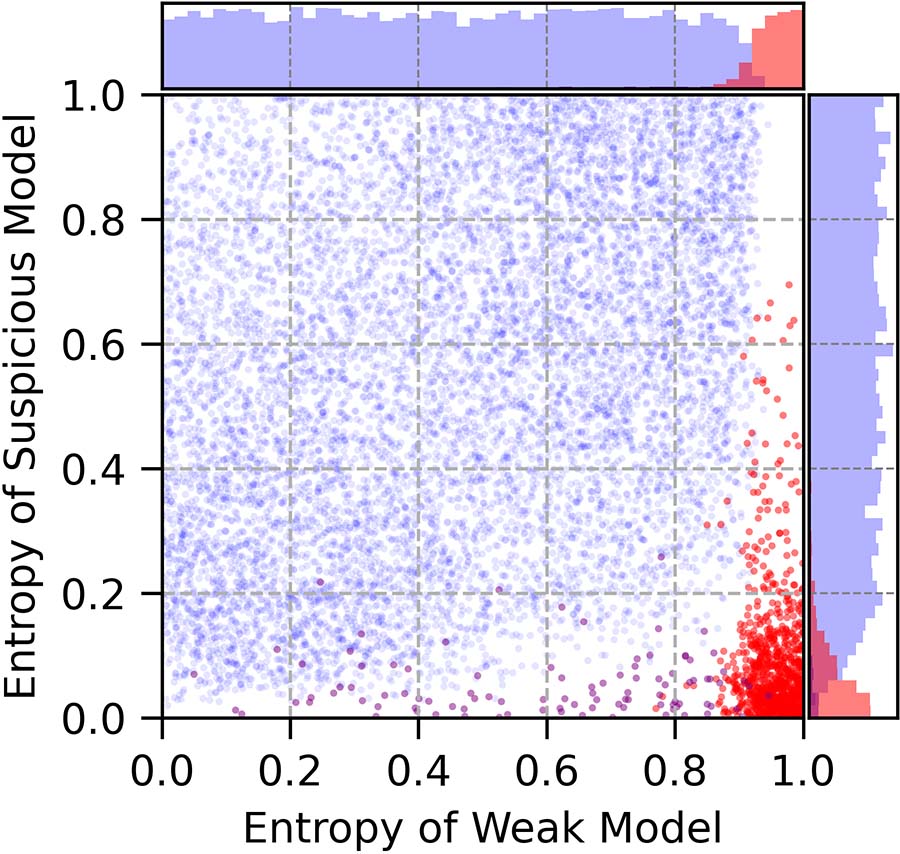}
    \caption{SSBA}
    \label{fig:ssba}
\end{subfigure}
\begin{subfigure}[b]{\subfigwidth}
    \centering
    \includegraphics[width=\linewidth]{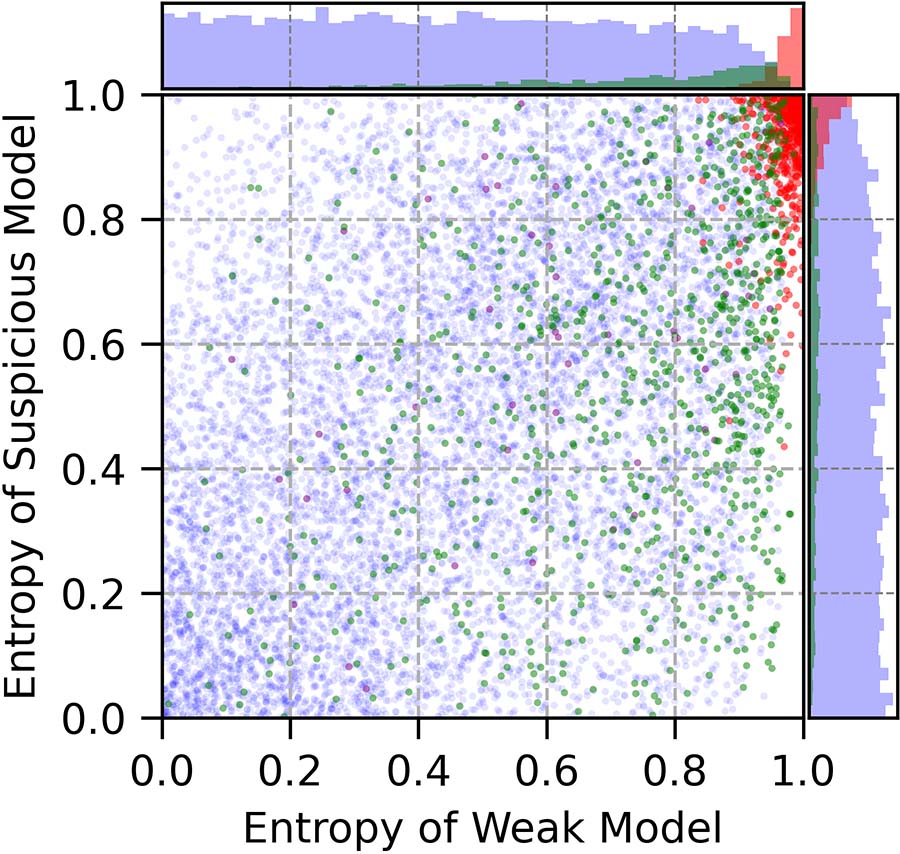}
    \caption{WaNet}
    \label{fig:wanet}
\end{subfigure}
\begin{subfigure}[b]{\subfigwidth}
    \centering
    \includegraphics[width=\linewidth]{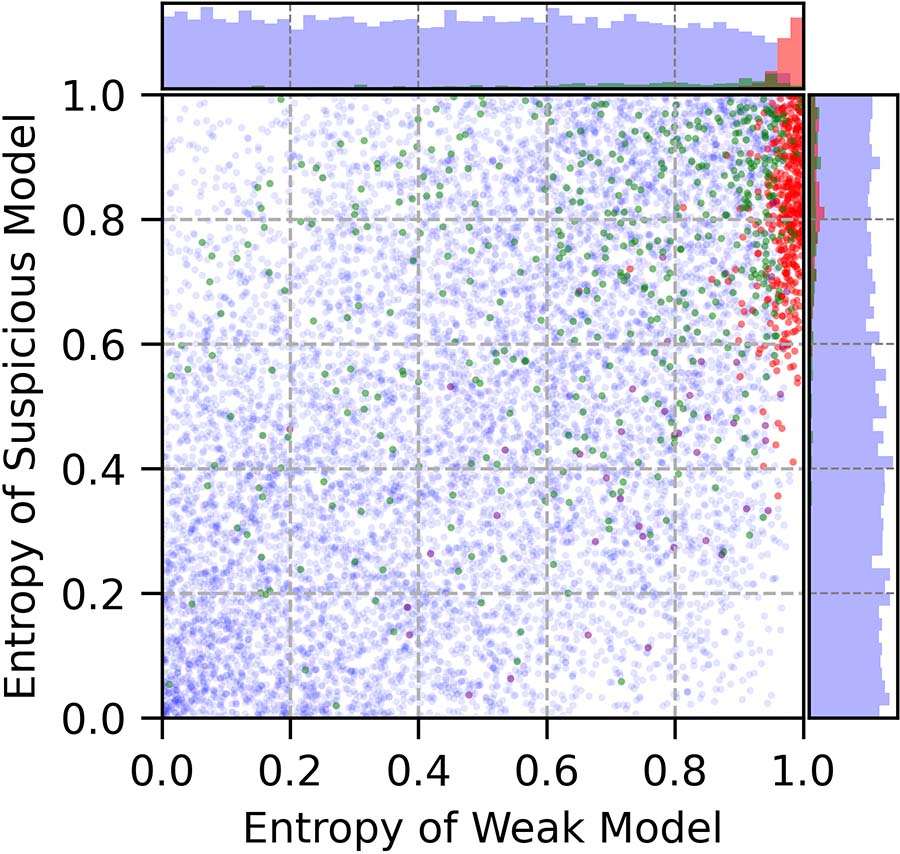}
    \caption{BPP}
    \label{fig:bpp}
\end{subfigure}
\begin{subfigure}[b]{\subfigwidth}
    \centering
    \includegraphics[width=\linewidth]{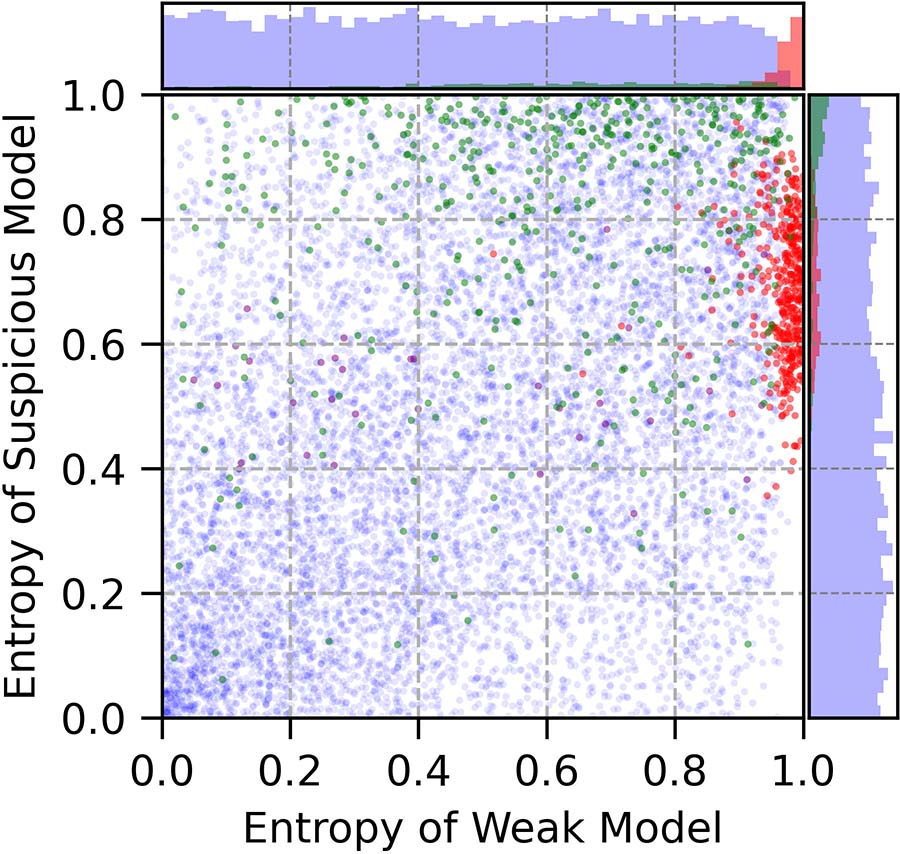}
    \caption{IAB}
    \label{fig:inputaware}
\end{subfigure}
\begin{subfigure}[b]{\subfigwidth}
    \centering
    \includegraphics[width=\linewidth]{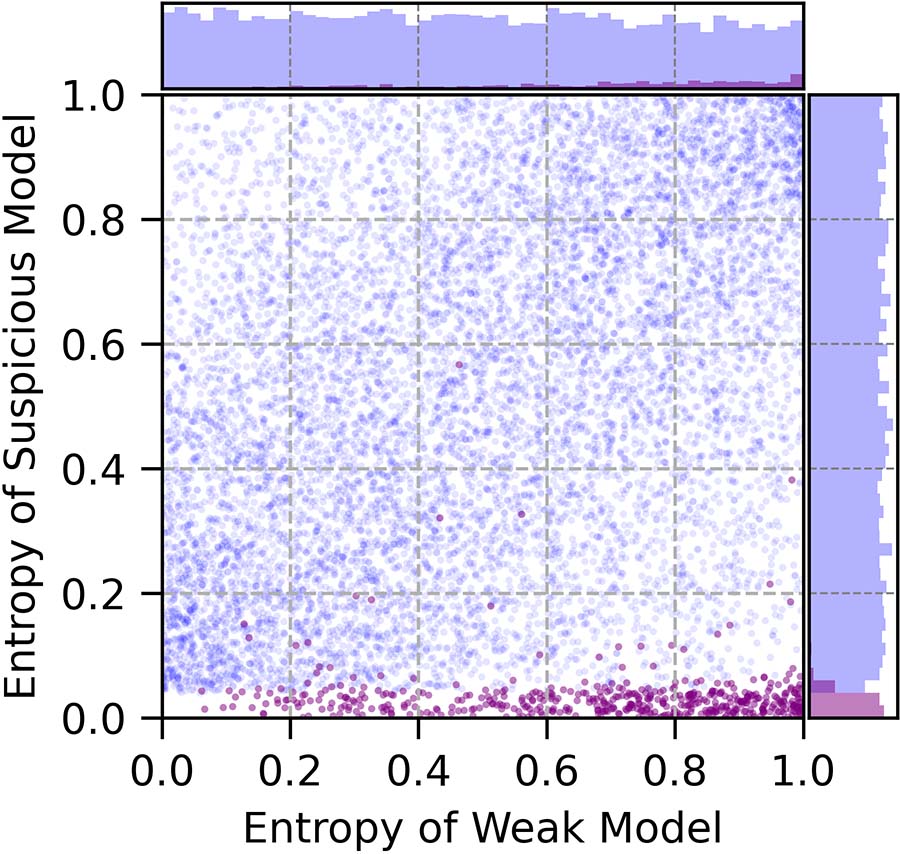}
    \caption{CTRL}
    \label{fig:ctrl}
\end{subfigure}
\begin{subfigure}[b]{\subfigwidth}
    \centering
    \includegraphics[width=\linewidth]{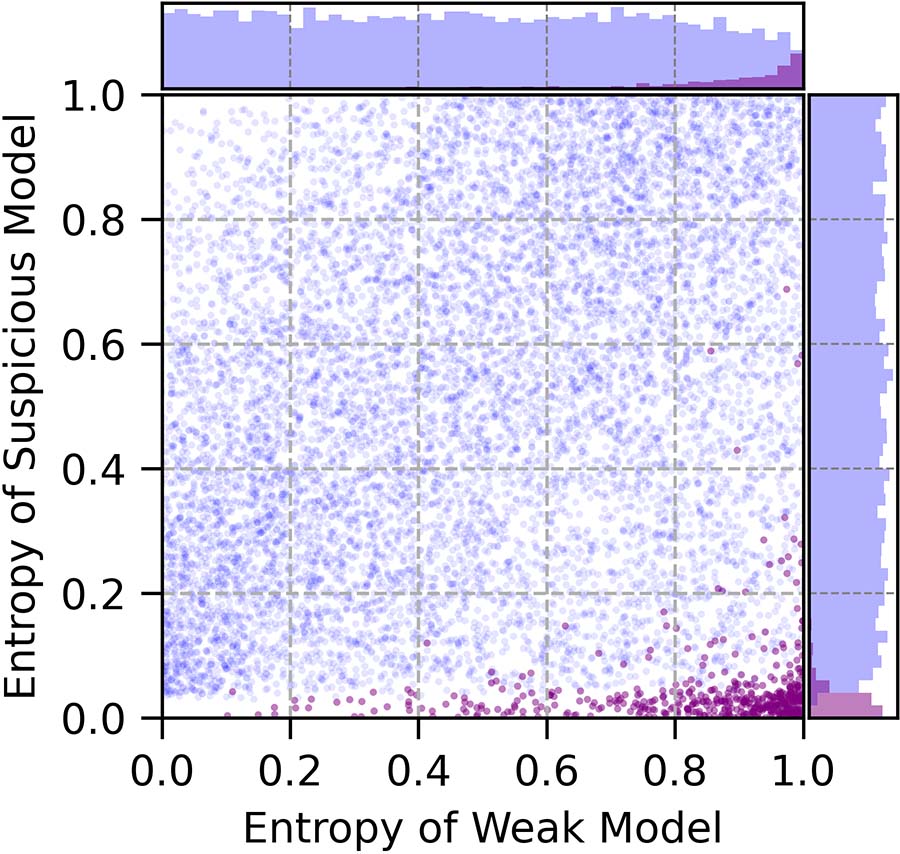}
    \caption{SIG}
    \label{fig:sig}
\end{subfigure}
\begin{subfigure}[b]{\subfigwidth}
    \centering
    \includegraphics[width=\linewidth]{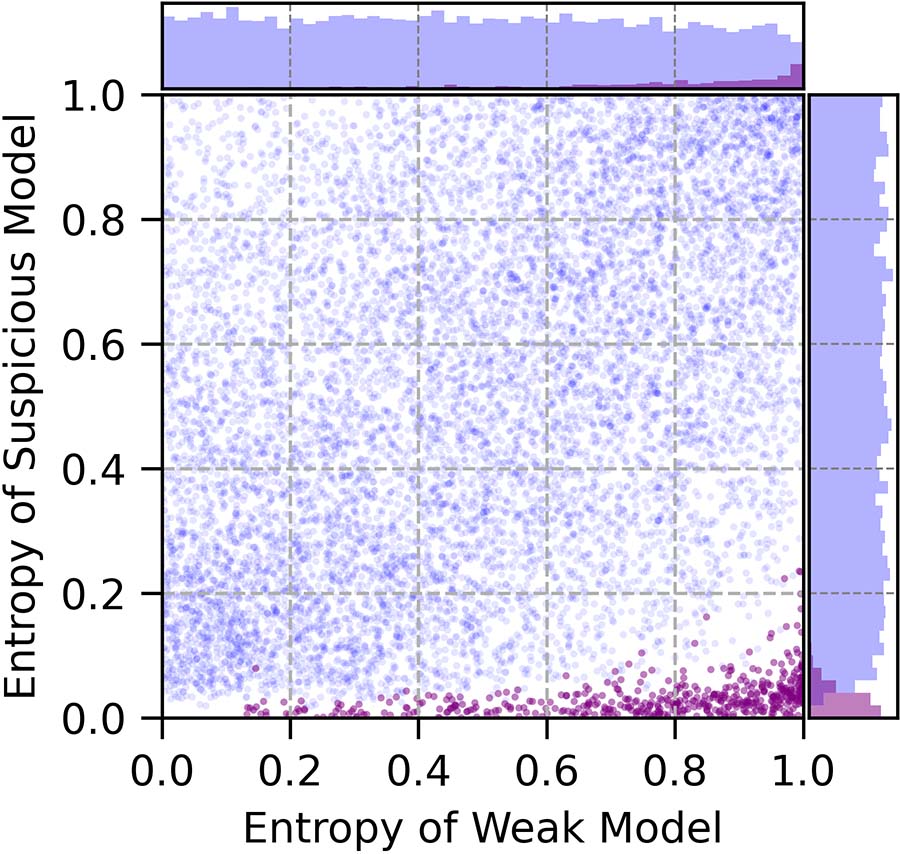}
    \caption{LC}
    \label{fig:lc}
\end{subfigure}
\begin{subfigure}[b]{\subfigwidth}
    \centering
    \includegraphics[width=\linewidth]{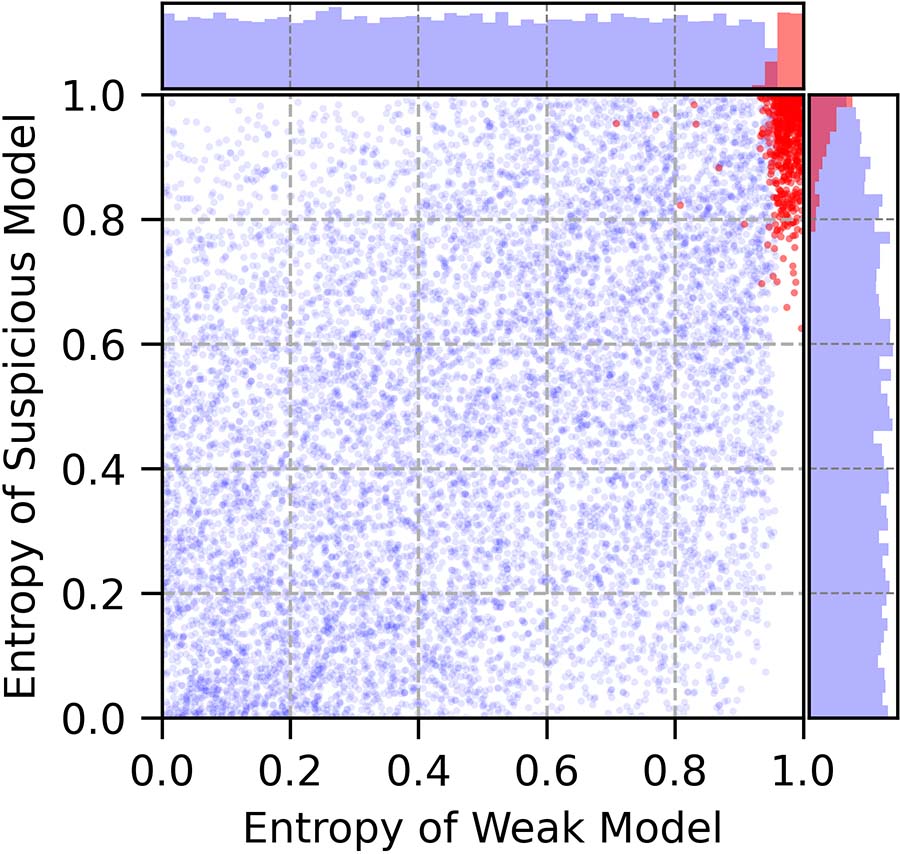}
    \caption{FLIP}
    \label{fig:flip}
\end{subfigure}
\begin{subfigure}[b]{\subfigwidth}
    \centering
    \includegraphics[width=\linewidth]{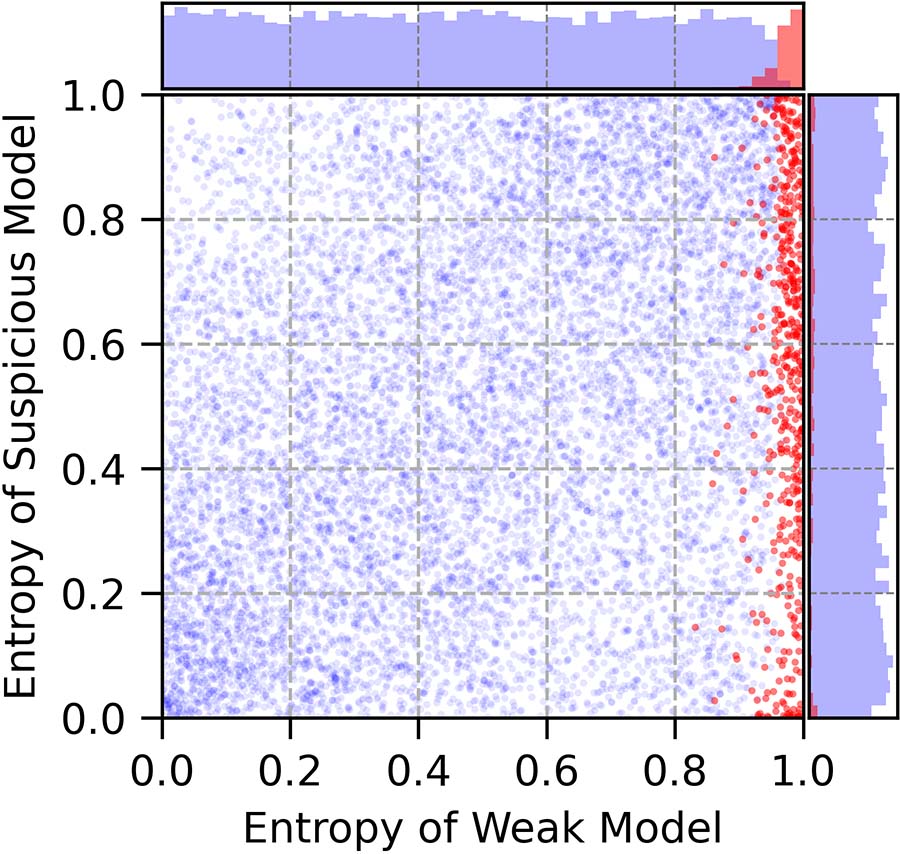}
    \caption{GCB}
    \label{fig:gcb}
\end{subfigure}
\begin{subfigure}[b]{\subfigwidth}
    \centering
    \includegraphics[width=\linewidth]{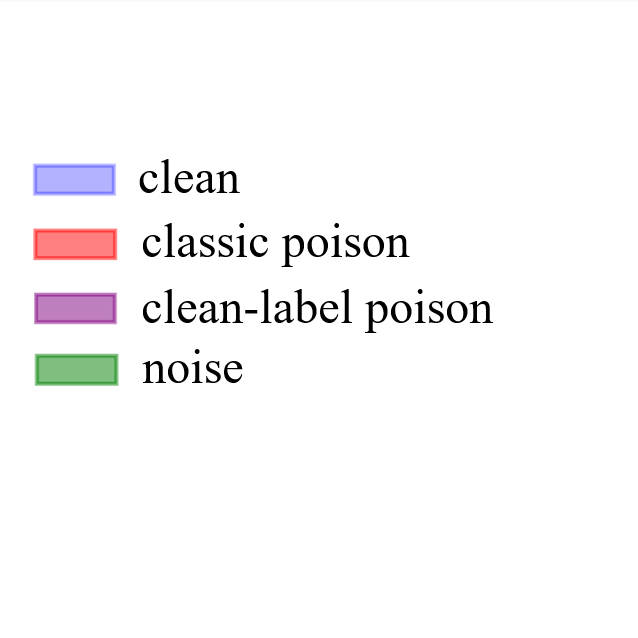}
\end{subfigure}

\caption{Entropy maps of various backdoor attack methods on the CIFAR-10 dataset. Each category of attack exhibits a distinct entropy distribution. Notably, SSBA, though classified as a dynamic backdoor attack, produces an entropy map resembling that of classic attacks. This similarity arises because SSBA does not introduce noise samples to prevent the victim network from learning trivial patterns. As a result, while the trigger appears dynamic, it behaves as a static pattern to the victim model.}
\label{fig:entropy_map}
\vskip -0.0in
\end{figure*}

\subsection{Experimental Results and Analysis}

Table \ref{tab:adaptive} summarizes the effectiveness of our CGD defense against adaptive attacks. The results demonstrate strong resistance to both strategies.

\begin{enumerate}
    \item \textbf{Adaptive-SIG Methods:} After applying our defense, the Attack Success Rate (ASR) drops significantly, remaining below 2.4\%. Although some weakly poisoned samples with minimal trigger strength exceed our detection threshold, most strongly triggered samples in the evaluated runs are identified and unlearned (Fig. \ref{fig:adasig}). This outcome indicates that while the attackers' efforts increase the stealthiness of the triggers, our defense remains robust against the more impactful threats in these experiments.

    \item \textbf{Feature Mixing Backdoor (FMB) Attack:} The ASR after defense is consistently low, under 1.2\% for both one-to-one and all-to-one scenarios. As shown in Fig. \ref{fig:fmb}, the CLIP model may exhibit uncertainty when predicting correct labels for mixed-feature images, but it also assigns high cross-entropy scores to the attackers' target labels. This dual uncertainty ensures that the poisoned samples are spotted by our defense mechanism, rendering the FMB attack ineffective.
\end{enumerate}

\subsubsection{\textbf{Analysis.}} The key to our defense's effectiveness lies in its dual examination of model outputs and cross-entropy measures. For the Adaptive-SIG methods, the randomization and reduced opacity introduce variability that can occasionally slip past entropy-based detection but fail to protect stronger triggers, which are crucial for a successful attack. In the case of the FMB attack, the inherent ambiguity of mixed features undermines the attacker's goal by preventing the model from confidently associating the poisoned samples with the target label without raising suspicion.

\section{More results on entropy map}
\label{ap:entropy_map}

The entropy maps in Fig. \ref{fig:entropy_map} reveal distinct patterns corresponding to four major backdoor attack categories: \textbf{Classical Backdoors}, \textbf{Dynamic Backdoors}, \textbf{Clean-label Backdoors}, and \textbf{Clean-image Backdoors}. Here, ``entropy'' refers to the percentile rank of the label-conditioned cross-entropy (negative log-likelihood) score defined in Section~\ref{subsec:meta_data_splitting}, rather than the Shannon entropy of the predictive distribution. For each attack type, we analyze this score for both the weak model and a suspicious model trained on the poisoned dataset.

\subsubsection{\textbf{Classical backdoors}} Classical backdoors, such as BadNets and Blend (Fig. \ref{fig:badnet} and \ref{fig:blended}), exhibit a characteristic pattern where poisoned samples (marked in red) cluster in the lower right corner of the entropy map. Because the horizontal axis is the weak-model score and the vertical axis is the suspicious-model score, this corresponds to high label-conditioned cross-entropy in the weak model and low cross-entropy in the suspicious model. The sharp clustering indicates that many classical poisoned samples can be separated using either label inconsistency under the weak model or trigger-induced confidence under the suspicious model.

\subsubsection{\textbf{Dynamic backdoors}} Dynamic backdoors, represented by methods such as \textbf{IAB} and \textbf{BPP} (Fig. \ref{fig:inputaware} and \ref{fig:bpp}), involve the addition of noise or randomized triggers to poison samples. The entropy map for dynamic backdoors shows a more dispersed pattern, with green points representing noisy samples. These samples occupy a wider entropy range in both the weak and suspicious models. For example, \textbf{IAB} introduces noise to prevent the network from learning trivial patterns, leading to a distribution that is challenging to distinguish from clean data. Interestingly, \textbf{SSBA} (Fig. \ref{fig:ssba}), despite being categorized as a dynamic backdoor, displays a pattern more akin to classical backdoors. This discrepancy occurs because SSBA does not introduce noise samples, resulting in a static trigger behavior that the victim model interprets as a fixed pattern. Hence, SSBA behaves more like a classical backdoor from an entropy perspective.

\subsubsection{\textbf{Clean-label backdoors}} Clean-label backdoors, such as \textbf{SIG}, \textbf{CTRL}, and \textbf{LC} (Fig. \ref{fig:sig}, \ref{fig:ctrl}, and \ref{fig:lc}), exhibit a unique entropy pattern distinguishable primarily through the suspicious model’s entropy values. Clean-label attacks rely on naturally occurring features, blending triggers with benign data, leading to entropy values that overlap significantly with clean samples. The entropy maps for these attacks show that poisoned samples (marked in red) are less separable in the weak model. However, the suspicious model’s entropy values for these samples tend to deviate from clean ones, allowing for the application of entropy-based detection methods. Notably, entropy-based defenses like Anti-Backdoor Learning (ABL) \cite{li2021anti} are effective against both clean-label and classical backdoors by leveraging these unique entropy distributions.

\subsubsection{\textbf{Clean-image Backdoors}} Clean-image backdoors, such as \textbf{FLIP} and \textbf{GCB} (Fig. \ref{fig:flip} and \ref{fig:gcb}), modify training labels without modifying the corresponding training images. Their label inconsistency produces high weak-model cross-entropy scores. FLIP also concentrates toward high suspicious-model scores, whereas GCB spans a broader range on the suspicious-model axis. Consequently, a suspicious-model loss alone may not reliably isolate these attacks, while the weak-model label-consistency signal used by CGD remains informative in the evaluated settings.

\section{Generalizability of Our Defense}
\label{ap:generalizability}

To evaluate the robustness and generalizability of our defense mechanism, we conducted experiments under different poison rates and model architectures. The default setup for our evaluations uses a PreActResNet18 model with a 5\% poison rate. However, we extend these tests to various scenarios to ensure that our defense is effective across a range of conditions.

\subsection{Effectiveness Across Poison Rates}
Table \ref{tab:pr} illustrates the attack performance under different poison rates, specifically at 1\% and 10\%. In both cases, we observe that our defense (CGD) significantly reduces the Attack Success Rate (ASR) while maintaining a high Clean Accuracy (CA). For instance, at a 1\% poison rate, the ASR for the BadNets attack is reduced from 73.8\% without defense to 2.7\% with our CGD defense. Similarly, at a 10\% poison rate, CGD reduces the ASR for most attacks to below 1\%, while the baseline ASR without defense remains above 90\% for most attacks. This demonstrates that our defense is resilient even as the poison rate increases, effectively mitigating attacks and preserving model accuracy.

\begin{table}
\centering
\caption{Attack performance under different poison rates.}
\label{tab:pr}
\vspace{-0.3cm}
\begin{small}
\begin{tblr}{
  width = \linewidth,
  rowsep = 0.8pt,
  colsep = 0.8pt,
  colspec = {Q[200]Q[83]Q[102]Q[88]Q[85]Q[92]Q[92]Q[88]Q[87]},
  cells = {c},
  cell{1}{2} = {c=4}{0.358\linewidth},
  cell{1}{6} = {c=4}{0.359\linewidth},
  cell{2}{2} = {c=2}{0.185\linewidth},
  cell{2}{4} = {c=2}{0.175\linewidth},
  cell{2}{6} = {c=2}{0.184\linewidth},
  cell{2}{8} = {c=2}{0.175\linewidth},
  cell{2}{1} = {r=2}{},
  hline{1,16} = {-}{0.10em},
  hline{2,4,15} = {-}{0.05em},
  hline{3} = {2-9}{0.03em},
}
Poison Rate & 1\%        &       &            &     & 10\%       &       &            &     \\
Attack $\downarrow$  & No Defense &       & CGD (ours) &     & No Defense &       & CGD (ours) &     \\
            & CA         & ASR   & CA         & ASR & CA         & ASR   & CA         & ASR \\
BadNets     & 93.2       & 73.8  & 92.6       & 2.7 & 91.8       & 93.8  & 92.3       & 0.0 \\
Blend       & 93.8       & 94.1  & 93.5       & 0.0 & 93.7       & 99.8  & 93.5       & 0.0 \\
WaNet       & 91.1       & 72.0  & 93.8       & 0.6 & 90.6       & 96.9  & 93.9       & 0.4 \\
BPP         & 91.4       & 99.2  & 93.9       & 0.1 & 91.4       & 99.2  & 93.9       & 0.1 \\
IAB         & 90.5       & 54.6  & 93.5       & 0.2 & 89.7       & 94.9  & 93.6       & 0.8 \\
SSBA        & 93.4       & 99.7  & 93.3       & 0.4 & 93.0       & 97.3  & 93.0       & 0.0 \\
CTRL        & 94.0       & 55.3  & 93.7       & 4.6 & 93.6       & 95.9  & 93.1       & 0.9 \\
SIG         & 93.7       & 80.4  & 93.5       & 0.0 & 84.6       & 98.0  & 84.4       & 0.0 \\
LC          & 93.5       & 68.3  & 93.2       & 2.8 & 84.4       & 99.8  & 84.4       & 0.0 \\
FLIP        & 93.2       & 98.1  & 93.3       & 0.1 & 85.4       & 99.7  & 91.5       & 0.0 \\
GCB         & 92.6       & 100.0 & 92.9       & 0.0 & 84.2       & 100.0 & 91.0       & 0.0 \\
Average     & 92.8       & 81.4 & 93.4       & 1.0 & 89.3       & 97.8 & 91.3       & 0.2 
\end{tblr}
\vspace{-0.3cm}
\end{small}
\end{table}

\begin{table}
\centering
\caption{Attack performance under different architectures.}
\label{tab:archi}
\vspace{-0.3cm}
\begin{small}
\begin{tblr}{
  width = \linewidth,
  rowsep = 0.8pt,
  colsep = 0.8pt,
  colspec = {Q[200]Q[83]Q[102]Q[88]Q[85]Q[92]Q[92]Q[88]Q[87]},
  cells = {c},
  cell{1}{2} = {c=4}{0.358\linewidth},
  cell{1}{6} = {c=4}{0.359\linewidth},
  cell{2}{2} = {c=2}{0.185\linewidth},
  cell{2}{4} = {c=2}{0.175\linewidth},
  cell{2}{6} = {c=2}{0.184\linewidth},
  cell{2}{8} = {c=2}{0.175\linewidth},
  cell{2}{1} = {r=2}{},
  hline{1,16} = {-}{0.08em},
  hline{2,4,15} = {-}{0.05em},
  hline{3} = {2-9}{0.03em},
}
Architecture & VGG16      &       &            &     & MobileNet V2 &      &            &     \\
Attack $\downarrow$ & No Defense &       & CGD (ours) &     & No Defense   &      & CGD (ours) &     \\
             & CA         & ASR   & CA         & ASR & CA           & ASR  & CA         & ASR \\
BadNets      & 89.5       & 95.0  & 90.0       & 0.0 & 82.1         & 90.5 & 81.5       & 1.7 \\
Blend        & 90.7       & 98.1  & 91.0       & 0.0 & 82.3         & 96.9 & 81.9       & 1.7 \\
WaNet        & 88.9       & 80.5  & 91.8       & 1.0 & 81.9         & 34.1 & 86.1       & 1.6 \\
BPP          & 88.6       & 99.4  & 91.8       & 1.0 & 82.6         & 94.6 & 86.2       & 1.6 \\
IAB          & 89.3       & 84.0  & 91.1       & 1.2 & 80.7         & 74.2 & 85.4       & 1.4 \\
SSBA         & 89.9       & 80.7  & 90.4       & 0.6 & 80.3         & 68.3 & 80.5       & 1.8 \\
CTRL         & 90.3       & 91.8  & 90.1       & 0.1 & 82.5         & 92.1 & 81.6       & 0.1 \\
SIG          & 90.4       & 97.2  & 90.3       & 0.0 & 81.3         & 98.7 & 80.0       & 0.0 \\
LC           & 90.3       & 87.8  & 90.0       & 0.0 & 82.6         & 96.4 & 81.6       & 0.1 \\
FLIP         & 88.4       & 98.2  & 90.6       & 0.2 & 81.4         & 93.0 & 81.3       & 0.1 \\
GCB          & 88.0       & 100.0 & 90.3       & 0.0 & 80.9         & 99.8 & 81.4       & 0.0 \\
Average      & 89.5       & 92.1 & 90.7       & 0.4 & 81.7         & 85.3 & 82.5       & 0.9 
\end{tblr}
\vspace{-0.3cm}
\end{small}
\end{table}

\subsection{Effectiveness Across Model Architectures}
To further demonstrate the generalizability of our defense, we evaluated it on different model architectures, including VGG16 and MobileNet V2, as shown in Table \ref{tab:archi}. Our CGD defense consistently reduces the ASR across both architectures. For instance, the ASR for the BadNets attack is reduced from 95.0\% to 0.0\% on VGG16 and from 90.5\% to 1.7\% on MobileNet V2. Similar trends are observed across other attack methods, demonstrating the robustness of our defense across different neural network architectures.

In summary, our defense method (CGD) demonstrates strong generalizability, maintaining its effectiveness across a range of poison rates and architectures. These results underscore the robustness of our approach in mitigating attacks across diverse settings, making it suitable for a variety of practical applications.

\section{More Results in Ablation Study}
We added an ablation here. Using only CLIP’s entropy mitigates most attacks (average ASR=13.8\%) but struggles with clean-label attacks (46.4\%). Model-only entropy excels there (ASR=1.3\%) but falters elsewhere (average ASR=25.1\%). Combining both yields the lowest average ASR (0.1\%), as shown in Table \ref{tab:abla_inf}. Related agent work studies when multi-to-single-agent skill distillation is beneficial~\cite{xu2026multi}.

\begin{table}
\centering
\caption{Ablation study for poisoned dataset splitting with entropy from different models.}
\label{tab:abla_inf}
\vspace{-0.3cm}
\begin{small}
\begin{tblr}{
  width  = 1.0\linewidth,
  rowsep = 0.5pt,
  colsep = 0.5pt,
  colspec = {Q[150]Q[60]Q[60]Q[60]Q[60]Q[80]Q[80]Q[80]Q[80]Q[60]Q[60]},
  cells  = {c},
  cell{1}{2}  = {c=2}{},
  cell{1}{4}  = {c=2}{},
  cell{1}{6}  = {c=2}{},
  cell{1}{8}  = {c=2}{},
  cell{1}{10} = {c=2}{},
  cell{3-5}{3} = {bg=customgreen},
  cell{3,5}{5} = {bg=customgreen},
  cell{4}{5} = {bg=customred},
  cell{4,5}{7} = {bg=customgreen},
  cell{3}{7} = {bg=customred},
  cell{3,5}{9} = {bg=customgreen},
  cell{4}{9} = {bg=customred},
  cell{3,5}{11} = {bg=customgreen},
  cell{4}{11} = {bg=customred},
  vline{2,10} = {1-5}{0.05em},
  hline{1,6}  = {-}{0.1em},
  hline{3}  = {-}{0.05em},
}
Category → & Classic & & Dynamic & & Clean-label & & Clean-image & & Average & \\ 
Method ↓   & CA  & ASR & CA  & ASR & CA  & ASR & CA  & ASR & CA  & ASR \\ 
CLIP only & 92.8 & 0.6 & 93.4 & 4.7 & 92.8 & 46.4 & 92.2 & 3.6 & 92.8 & 13.8 \\ 
Victim only & 92.8 & 1.0 & 92.7 & 57.7 & 93.3 & 1.3 & 90.4 & 38.7 & 92.3 & 25.1 \\ 
Both & 93.0 & 0.0 & 93.5 & 0.3 & 93.4 & 0.0 & 92.4 & 0.2 & 93.1 & 0.1 \\ 
\end{tblr}
\vspace{-0.0cm}
\end{small}
\end{table}

\section{Use CLIP Variants in Domain-Specific Tasks}

CGD’s reliance on a CLIP-like model is practical because defenders can use domain-specific CLIP variants for specific tasks. Across 17 datasets tested, CGD succeeded in 16, showcasing its broad applicability. For the sole exception, SVHN (vanilla CLIP accuracy: 13.4\%, near-random guess for 10-class dataset), switching to LAION-2B CLIP variant reduced the ASR to 2.6\% under the BPP attack. In agentic systems, contextual records should likewise not be conflated with true memory~\cite{xu2026contextual}.

In other domain-specific environments like medical imaging, CLIP models such as PubMed \cite{pubmedclip} (62.7\% accuracy on PCAM) achieved an ASR of 6.1\%, as shown in Table \ref{tab:clip_bpp}. These results demonstrate that with advanced domain-specific CLIP variants (e.g. PubMedCLIP \cite{pubmedclip} in Medicine, BioCLIP \cite{stevens2024bioclip} in Biology, GeomCLIP \cite{xiao2024geomclipcontrastivegeometrytextpretraining} in Chemistry), defenders can select high-performing CLIP variants on the poisoned dataset to ensure CGD’s effectiveness. CLIP’s public availability and ongoing advancements in domain-specific models further reduce feasibility concerns.

\begin{table}[t]
\centering
\caption{Performance of different CLIP pre-training weights on \textsc{SVHN} and \textsc{PCAM}.}
\vspace{-0.2cm}
\label{tab:clip_bpp}
\begin{small}
\begin{tblr}{
  width = 0.9\linewidth,
  rowsep = 0.2pt,
  colsep = 0.8pt,
  colspec = {Q[238]Q[254]Q[192]Q[113]Q[113]},
  cells  = {c},
  cell{1}{2} = {r=2}{},
  cell{1}{3} = {r=2}{},
  cell{1}{4} = {c=2}{},
  cell{3}{1} = {r=3}{},
  cell{6}{1} = {r=3}{},
  hline{1,9}  = {-}{0.1em},
  hline{3,6}  = {-}{0.05em},
}
Task→                  & CLIP Weight & CLIP Acc. & BPP Attack &      \\
Dataset↓               &             &           & CA~        & ASR  \\
{SVHN  \\(10 classes)} & Vanilla     & 13.4      & 94.3       & 100  \\
                       & LAION-400M \cite{schuhmann2022laion}  & 27.7      & 94.8       & 15.2 \\
                       & LAION-2B \cite{schuhmann2022laion}    & 38.8      & 95.6       & 2.6  \\
{PCAM  \\(2 classes)}  & Vanilla     & 52.3      & 82.1       & 69.5 \\
                       & BioMed \cite{zhang2025biomedclipmultimodalbiomedicalfoundation}      & 60.1      & 83.8       & 15.8 \\
                       & PubMed \cite{pubmedclip}      & 62.7      & 85.4       & 6.1  
\end{tblr}
\end{small}
\vspace{-0.1cm}
\end{table}

\begin{table*}
\centering
\caption{Clean-data-based defenses (5\%) vs poisoned-data-based defenses (100\%). In this case, poisoned-data-based defenses can get better defense performance than clean-data-based defenses with the help of ``SC'' (Split Clean data).}

\vspace{-0.3cm}
\label{tab:5percent_clean}
\begin{tblr}{
  width = 1.0\linewidth,
  rowsep = 0.2pt,
  colsep = 0.1pt,
  colspec = {Q[94]Q[71]Q[46]Q[46]Q[46]Q[46]Q[46]Q[46]Q[46]Q[54]Q[46]Q[46]Q[46]Q[46]Q[48]Q[48]Q[46]Q[46]},
  cells = {c},
  cell{1}{1} = {c=2}{},
  cell{1}{3} = {c=8}{},
  cell{1}{11} = {c=8}{},
  cell{2}{1} = {c=2}{},
  cell{2}{3} = {c=2}{},
  cell{2}{5} = {c=2}{},
  cell{2}{7} = {c=2}{},
  cell{2}{9} = {c=2}{},
  cell{2}{11} = {c=2}{},
  cell{2}{13} = {c=2}{},
  cell{2}{15} = {c=2}{},
  cell{2}{17} = {c=2}{},
  cell{3}{1} = {c=2}{},
  cell{4}{1} = {r=2}{},
  cell{6}{1} = {r=4}{},
  cell{10}{1} = {r=3}{},
  cell{13}{1} = {r=2}{},
  cell{15}{1} = {c=2}{},
  cell{4,6,9,13}{4} = {bg = customgreen},
cell{5,7,8,10,11,12,14,15}{4} = {bg = customred},
cell{4,6-10,13-15}{6} = {bg = customgreen},
cell{5,11,12}{6} = {bg = customred},
cell{4,6,9,11-15}{8} = {bg = customgreen},
cell{5,7,8,10}{8} = {bg = customred},
cell{4,6,7,13}{10} = {bg = customgreen},
cell{5,8-12,14,15}{10} = {bg = customred},
cell{4,6,8,9,12-14}{12} = {bg = customgreen},
cell{5,7,10,11,15}{12} = {bg = customred},
cell{4,6-15}{14} = {bg = customgreen},
cell{5}{14} = {bg = customred},
cell{4,6-9,11-13,15}{16} = {bg = customgreen},
cell{5,10,14}{16} = {bg = customred},
cell{4-13,15}{18} = {bg = customgreen},
cell{14}{18} = {bg = customred},
  vline{3,11} = {-}{0.05em},
  hline{1,16} = {-}{0.1em},
  hline{4,6,10,13,15} = {-}{0.05em},
  hline{3} = {3,5,7,9,11,13,15,17,19}{l},
  hline{3} = {4,6,8,10,12,14,16,18,20}{r},
}
\textbf{Requirement →}     &         & \textbf{5\% Clean Training Data} &      &               &      &                 &      &               &       & \textbf{100\% Poisoned Training Data } &      &                 &      &                   &      &                 &      \\
\textbf{Defense →}         &         & \textbf{FP \cite{liu2018fp}}                     &      & \textbf{ANP \cite{wu2021anp}} &      & \textbf{I-BAU \cite{zeng2021ibau}} &      & \textbf{MCR \cite{zhao2020mcr}} &       & \textbf{FP-SC}                         &      & \textbf{ANP-SC} &      & \textbf{I-BAU-SC} &      & \textbf{MCR-SC} &      \\
\textbf{Attack ↓ }         &         & CA                               & ASR  & CA            & ASR  & CA              & ASR  & CA            & ASR   & CA                                     & ASR  & CA              & ASR  & CA                & ASR  & CA              & ASR  \\
{Classic \\Backdoor }      & BadNets & 92.3                             & 2.1  & 85.6          & 0.0  & 86.2            & 0.8  & 92.1          & 2.4   & 93.3                                   & 1.0  & 85.4            & 0.0  & 90.1              & 1.5  & 91.2            & 1.0  \\
                           & Blend   & 91.9                             & 41.4 & 85.0          & 42.7 & 86.5            & 32.3 & 93.1          & 99.8  & 93.3                                   & 35.3 & 87.9            & 37.6 & 91.3              & 49.9 & 90.7            & 1.9  \\
{Dynamic~\\Backdoor   }    & WaNet   & 93.2                             & 5.4  & 82.6          & 0.6  & 90.3            & 3.8  & 93.4          & 12.9  & 94.5                                   & 2.1  & 85.5            & 0.1  & 91.0              & 5.1  & 92.1            & 1.1  \\
                           & BPP     & 93.2                             & 66.9 & 82.7          & 0.6  & 90.9            & 91.6 & 93.4          & 2.0   & 94.3                                   & 89.0 & 87.3            & 0.5  & 92.7              & 13.1 & 91.7            & 2.2  \\
                           & IAB     & 92.9                             & 36.8 & 85.5          & 1.3  & 90.8            & 24.2 & 93.0          & 53.6  & 94.1                                   & 2.1  & 90.6            & 0.3  & 92.3              & 2.4  & 91.7            & 2.5  \\
                           & SSBA    & 92.0                             & 14.5 & 87.3          & 0.1  & 90.0            & 1.5  & 92.4          & 64.2  & 92.9                                   & 7.7  & 86.1            & 0.0  & 91.2              & 4.3  & 90.5            & 1.0  \\
{Clean-Label\\Backdoor  }  & CTRL    & 92.5                             & 98.5 & 93.0          & 3.8  & 89.5            & 25.4 & 93.5          & 99.3  & 93.1                                   & 78.5 & 86.1            & 1.9  & 90.2              & 41.2 & 90.6            & 2.5  \\
                           & SIG     & 92.7                             & 40.9 & 86.3          & 22.5 & 87.7            & 17.5 & 93.4          & 93.9  & 93.2                                   & 73.9 & 86.3            & 12.1 & 90.1              & 10.6 & 90.3            & 13.2 \\
                           & LC      & 92.1                             & 34.7 & 87.6          & 51.1 & 88.9            & 5.3  & 93.6          & 100.0 & 93.1                                   & 19.3 & 85.5            & 16.7 & 92.1              & 1.6  & 90.4            & 1.4  \\
{Clean-Image\\Backdoor   } & FLIP    & 91.9                             & 0.1  & 91.7          & 0.1  & 87.9            & 0.3  & 91.9          & 0.3   & 93.1                                   & 12.7 & 88.0            & 0.0  & 90.1              & 0.3  & 90.6            & 0.3  \\
                           & GCB     & 92.3                             & 35.4 & 81.2          & 0.0  & 90.9            & 12.8 & 91.2          & 90.0  & 92.4                                   & 19.0 & 81.4            & 0.0  & 90.7              & 86.4 & 90.3            & 92.4 \\
Average                    &         & 92.5                             & 34.2 & 86.2          & 11.2 & 89.1            & 19.6 & 92.8          & 56.2  & 93.4                                   & 31.0 & 86.4            & 6.3  & 91.1              & 19.7 & 90.9            & 10.9 
\end{tblr}
\end{table*}

\begin{table*}
\centering
\caption{Clean-data-based defenses (5\%) vs poisoned-data-based defenses (10\%). In this case, poisoned-data-based defenses can get comparable defense performance to clean-data-based defenses with the help of ``SC'' (Split Clean data).}
\vspace{-0.3cm}
\label{tab:10percent_poison}
\begin{tblr}{
  width = 1.0\linewidth,
  rowsep = 0.2pt,
  colsep = 0.1pt,
  colspec = {Q[94]Q[71]Q[46]Q[46]Q[46]Q[46]Q[46]Q[46]Q[46]Q[54]Q[46]Q[46]Q[46]Q[46]Q[48]Q[48]Q[46]Q[46]},
  cells = {c},
  cell{1}{1} = {c=2}{},
  cell{1}{3} = {c=8}{},
  cell{1}{11} = {c=8}{},
  cell{2}{1} = {c=2}{},
  cell{2}{3} = {c=2}{},
  cell{2}{5} = {c=2}{},
  cell{2}{7} = {c=2}{},
  cell{2}{9} = {c=2}{},
  cell{2}{11} = {c=2}{},
  cell{2}{13} = {c=2}{},
  cell{2}{15} = {c=2}{},
  cell{2}{17} = {c=2}{},
  cell{3}{1} = {c=2}{},
  cell{4}{1} = {r=2}{},
  cell{6}{1} = {r=4}{},
  cell{10}{1} = {r=3}{},
  cell{13}{1} = {r=2}{},
  cell{15}{1} = {c=2}{},
  cell{4,6,9,13}{4} = {bg = customgreen},
cell{5,7,8,10,11,12,14,15}{4} = {bg = customred},
cell{4,6-10,13-15}{6} = {bg = customgreen},
cell{5,11,12}{6} = {bg = customred},
cell{4,6,9,11-15}{8} = {bg = customgreen},
cell{5,7,8,10}{8} = {bg = customred},
cell{4,6,7,13}{10} = {bg = customgreen},
cell{5,8-12,14,15}{10} = {bg = customred},
cell{4,6,8,9,13}{12} = {bg = customgreen},
cell{5,7,10-12,14,15}{12} = {bg = customred},
cell{4,6-11,13-15}{14} = {bg = customgreen},
cell{5,12}{14} = {bg = customred},
cell{4,7-13,15}{16} = {bg = customgreen},
cell{5,6,14}{16} = {bg = customred},
cell{6-8}{18} = {bg = customgreen},
cell{4,5,9-15}{18} = {bg = customred},
  vline{3,11} = {-}{0.05em},
  hline{1,16} = {-}{0.1em},
  hline{4,6,10,13,15} = {-}{0.05em},
  hline{3} = {3,5,7,9,11,13,15,17,19}{l},
  hline{3} = {4,6,8,10,12,14,16,18,20}{r},
}
\textbf{Requirement →}     &         & \textbf{5\% Clean Training Data} &      &               &      &                 &      &               &       & \textbf{10\% Poisoned Training Data } &      &                 &      &                   &      &                 &      \\
\textbf{Defense →}         &         & \textbf{FP \cite{liu2018fp}}                     &      & \textbf{ANP \cite{wu2021anp}} &      & \textbf{I-BAU \cite{zeng2021ibau}} &      & \textbf{MCR \cite{zhao2020mcr}} &       & \textbf{FP-SC}                         &      & \textbf{ANP-SC} &      & \textbf{I-BAU-SC} &      & \textbf{MCR-SC} &      \\
\textbf{Attack ↓ }         &         & CA                               & ASR  & CA            & ASR  & CA              & ASR  & CA            & ASR   & CA                                     & ASR  & CA              & ASR  & CA                & ASR  & CA              & ASR  \\
{Classic \\Backdoor }      & BadNets & 92.3                             & 2.1  & 85.6          & 0.0  & 86.2            & 0.8  & 92.1          & 2.4   & 92.5                                  & 2.7  & 85.0            & 0.0  & 89.5              & 1.2  & 92.8            & 34.1  \\
                           & Blend   & 91.9                             & 41.4 & 85.0          & 42.7 & 86.5            & 32.3 & 93.1          & 99.8  & 92.5                                  & 37.6 & 88.4            & 42.5 & 89.0              & 59.4 & 93.5            & 99.3  \\
{Dynamic~\\Backdoor   }    & WaNet   & 93.2                             & 5.4  & 82.6          & 0.6  & 90.3            & 3.8  & 93.4          & 12.9  & 89.6                                  & 2.1  & 82.2            & 0.1  & 91.8              & 22.1 & 93.1            & 0.6   \\
                           & BPP     & 93.2                             & 66.9 & 82.7          & 0.6  & 90.9            & 91.6 & 93.4          & 2.0   & 93.2                                  & 90.4 & 88.1            & 0.8  & 91.7              & 9.0  & 92.9            & 2.8   \\
                           & IAB     & 92.9                             & 36.8 & 85.5          & 1.3  & 90.8            & 24.2 & 93.0          & 53.6  & 93.3                                  & 1.9  & 83.7            & 0.0  & 91.8              & 2.9  & 92.9            & 1.2   \\
                           & SSBA    & 92.0                             & 14.5 & 87.3          & 0.1  & 90.0            & 1.5  & 92.4          & 64.2  & 88.6                                  & 16.1 & 86.0            & 0.0  & 90.2              & 10.9 & 92.5            & 51.5  \\
{Clean-Label\\Backdoor  }  & CTRL    & 92.5                             & 98.5 & 93.0          & 3.8  & 89.5            & 25.4 & 93.5          & 99.3  & 91.1                                  & 95.3 & 93.2            & 2.3  & 90.5              & 6.5  & 93.6            & 99.6  \\
                           & SIG     & 92.7                             & 40.9 & 86.3          & 22.5 & 87.7            & 17.5 & 93.4          & 93.9  & 92.6                                  & 55.7 & 88.1            & 7.3  & 88.6              & 7.6  & 93.5            & 97.8  \\
                           & LC      & 92.1                             & 34.7 & 87.6          & 51.1 & 88.9            & 5.3  & 93.6          & 100.0 & 92.1                                  & 47.8 & 89.3            & 56.6 & 88.5              & 7.1  & 92.9            & 100.0 \\
{Clean-Image\\Backdoor   } & FLIP    & 91.9                             & 0.1  & 91.7          & 0.1  & 87.9            & 0.3  & 91.9          & 0.3   & 92.1                                  & 1.9  & 90.5            & 0.0  & 88.2              & 3.4  & 92.5            & 76.7  \\
                           & GCB     & 92.3                             & 35.4 & 81.2          & 0.0  & 90.9            & 12.8 & 91.2          & 90.0  & 92.1                                  & 59.5 & 84.5            & 3.2  & 89.3              & 87.6 & 93.2            & 89.1  \\
Average                    &         & 92.5                             & 34.2 & 86.2          & 11.2 & 89.1            & 19.6 & 92.8          & 56.2  & 91.8                                  & 37.4 & 87.2            & 10.3 & 89.9              & 19.8 & 93.0            & 59.3  
\end{tblr}
\end{table*}

\section{More Results on Defenses Based on Clean Data}
\label{ap:clean_data_based}

As mentioned in the introduction, we introduce a novel insight that enables defenses traditionally dependent on clean data to operate effectively on poisoned data by leveraging CGD to filter a clean subset. This concept opens up the possibility for adapting clean-data defenses for poisoned environments, offering a practical solution when clean data is unavailable. In this section, we evaluate clean-data based defenses under three conditions: 5\% clean training dataset, whole (100\%) poisoned training dataset, and 10\% poisoned training dataset.

Table \ref{tab:5percent_clean} presents the results of clean-data based defenses with 5\% clean training data available. In this setting, the total number of successful defenses (defined as defenses achieving an Attack Success Rate (ASR) of 20\% or lower, highlighted in green) is 23.

Table \ref{tab:5percent_clean} shows the results for clean-data based defenses when a clean subset is split from the whole (100\%) poisoned training dataset. With access to the full poisoned dataset, the defense performance improves significantly, with the number of successful defenses increasing to 35, marking a 50\% improvement over the 5\% clean data setting. This indicates that having access to the full poisoned dataset allows the defenses to operate more effectively, even in the absence of originally clean data.

Table \ref{tab:10percent_poison} provides results for clean-data based defenses with a 10\% poisoned training dataset. Even in this setting, where only a small portion of the dataset is clean, the total number of successful defenses increases from 23 (in the 5\% clean data setting) to 25. This demonstrates that a 10\% poisoned dataset can yield comparable performance to that obtained with 5\% clean data, suggesting that even minimal clean data availability can significantly bolster defense effectiveness.

In summary, our findings suggest that accessing the full (100\%) poisoned dataset enables all tested defense methods to achieve better performance. The total number of successful defenses (ASR $\leq$ 20\%) increased from 23 to 35, representing a 50\% improvement over the original method that used only 5\% clean data. Even with a 10\% poisoned dataset, the number of successful defenses increased from 23 to 25, indicating that 10\% poisoned data achieves a comparable performance to 5\% clean data. This demonstrates the potential of clean-data-based defenses to perform effectively in poisoned environments, even with limited clean data availability.

\end{document}